\documentclass[eqsecnum,superscriptaddress,nofootinbib,preprint,showpacs,11pt]{revtex4}

\usepackage{amsmath} 
\usepackage{amssymb} 
\usepackage{bbm} 
\usepackage{graphicx} 

\newcommand{\abs}[1]{{\left|{#1}\right|}} 
\newcommand{\ket}[1]{\vert{#1}\rangle} 
\newcommand{\bra}[1]{\langle{#1}\vert} 

\newcommand{\secref}[1]{Sec.~\ref{#1}}
\newcommand{\eqnref}[1]{(\ref{#1})}
\newcommand{\figref}[1]{Fig.~\ref{#1}}
\newcommand{\footref}[1]{Footnote~\ref{#1}}
\newcommand{\appref}[1]{Appendix~\ref{#1}}

\begin{document}

\title{Response of the Unruh-DeWitt detector in flat spacetime with a compact dimension}

\author{Dah-Wei Chiou}
\email{dwchiou@gmail.com}
\affiliation{Department of Physics, National Sun Yat-sen University, Kaohsiung 80424, Taiwan}
\affiliation{Center for Condensed Matter Sciences, National Taiwan University, Taipei 10617, Taiwan}


\begin{abstract}
In a flat spacetime with one spatial dimension compactified, inertial reference frames are not all equivalent, but there are the preferred ones. This paper investigates the nonequivalence of inertial frames and also that of uniformly accelerated frames in connection with the response of the Unruh-DeWitt detector coupled to a massless scalar field. The detector's transition rates of both excitation and de-excitation are studied in depth for three different cases: (i) the detector moving at an arbitrary constant velocity, (ii) moving with a constant acceleration in the compact direction, and (iii) moving with a constant acceleration in noncompact directions. The instantaneous transition rate in relation to the switching function is also taken into account.
\end{abstract}


\maketitle

\section{Motivations and overview}
The twin paradox is a well-known puzzle in the theory of relativity: one of the two identical twins travels on a high-speed spacecraft away from Earth and then turns around and comes back, while the other stays on Earth. The puzzle arises because each twin apparently sees the other as moving, and therefore time dilation seems to suggest that, paradoxically, the twins should find each other less aged by the time when they meet. As the paradox is resolved within the standard theory of relativity, it turns out the traveling twin is less aged than the earthbound sibling, not the other way around.

There have been various explanations of the twin paradox, all recognizing the crucial fact that the symmetry between the two twins is in fact illusory. The earthbound twin is in the same inertial (rest) frame all the time, while the traveling twin undergoes two different (outbound and inbound) initial frames throughout the journey. The frame switch upon the traveling twin is essentially the reason for the aging difference. The switch of initial frames implies that the traveling twin must experience acceleration during the period of turnaround, which can also be used to account for his slowed aging in terms of gravitational time dilation (although it is often argued that acceleration \textit{per se} plays no direct role). For more discussions on the twin paradox, see \cite{Debs:1996} and references therein.

The puzzle strikes back again when we consider the twin paradox in a flat spacetime with one spatial dimension compactified. If the traveling twin moves at a constant velocity in the compact direction, his frame remains inertial for the entire journey, yet the topology allows him to meet the earthbound twin after he circumnavigates the compact dimension. As the traveling twin undergoes no frame switch at all, the standard explanation of the aging difference no longer works. The resolution to the puzzle lies in the fact that compactifying a spatial dimension breaks the global Lorentz invariance. As a consequence, there is now a class of \emph{preferred} inertial reference frames, namely, those at rest in the compact direction \cite{Brans:1973,Peters:1983as,Dray:1990,Barrow:2001rj,Uzan:2000wp}.

As inertial frames are not all equivalent now, an observer, in principle, can experimentally determine the frame's moving velocity in the compact direction with respect to the preferred frame. This can be done by performing a ``global'' experiment: sending two light beams in opposite directions along the compact dimension and measuring the arrival time of both signals when they come back. The frame's moving velocity relative to the preferred frame can be inferred from the time delay between the two arriving signals \cite{Brans:1973,Peters:1983as}. On the other hand, a ``local'' experiment (as performed \emph{within} the comoving frame) is also possible. For instance, as one spatial dimension is compactified, the form of the electrostatic field of a point charge is deviated from $1/r^2$. Measuring the deviation can also determine the frame's moving velocity relative to the preferred frame \cite{Bansal:2005ue}.

It is instructive to look for other kinds of local experiments, as they will teach us to what extent the initial reference frames are inequivalent. Particularly, as the velocity relative to the preferred frame bears an absolute meaning now (and in a sense analogous to acceleration in the Minkowski spacetime), it is suggestive that the Unruh-DeWitt detector moving at a constant velocity might register signals revealing its velocity. Recently, it was shown that, in the Minkowski spacetime, coupled to a massless scalar field in the \emph{polymer quantization} (which implements some features of the microscopic discreteness in loop quantum gravity) \cite{Hossain:2010eb}, even the Unruh-DeWitt detector moving at constant velocity detects nontrivial radiation \cite{Kajuri:2015oza}. This is essentially because the Lorentz invariance is violated in the UV scale by the microscopic discreteness. In our case of flat spacetime with a compact dimension, as the Lorentz invariance is violated in the IR scale by the large length of the compact dimension, it is curious to know whether the Unruh-DeWitt detector moving at a constant velocity also detect nontrivial signals.

This paper investigates the response of the Unruh-DeWitt detector coupled to a massless scalar field in a flat spacetime with one spatial dimension compactified. It turns out, when the Unruh-DeWitt detector moves at a constant velocity, the detector's equilibrium transition rate of de-excitation depends on the moving velocity in the compact direction as well as the size of the compact dimension, implying that the nonequivalence of inertial frames is discernable by the response of the Unruh-DeWitt detector. The equilibrium transition rate of excitation, on the other hand, remains zero as in the ordinary (uncompactified) Minkowski spacetime. If one is able to switch the detector on and off at will and measure the instantaneous (nonequilibrium) transition rate accordingly, then the rates of both excitation and de-excitation are nonzero and can be used to infer the velocity in the compact direction and the size of compact dimension.

Furthermore, we also study the response of the Unruh-DeWitt detector moving with a constant acceleration both in the compact direction and in noncompact directions. When the detector accelerates in the compact direction, the detector's response never equilibrates and thus one can only make sense of the instantaneous transition rates, which, of both excitation and de-excitation, depend ont only on the detector's acceleration and the size of the compact dimension but also on the time when the observation is performed and the instantaneous velocity in the compact direction at the moment when the detector is turned on. On the other hand, when the detector accelerates in noncompact direction, the response is in equilibrium with the field. While the equilibrium transition rate of excitation remains to be the same celebrated form of thermal radiation as that in the ordinary Minkowski spacetime, the equilibrium transition rate of de-excitation exhibits an extra correction dependent on the size of the compact dimension.

This paper is organized as follows. It begins in \secref{sec:Unruh-DeWitt detector} with a brief review on the Unruh-DeWitt detector, mainly based on \cite{Birrell:1982ix}. The Wightman function is then derived in \secref{sec:Wightman function} for the flat spacetime with a compact dimension. In Secs.~\ref{sec:with switching} and \ref{sec:sharp switching limit}, the effects of a switching function and the sharp switching limit are carefully studied by following the treatments in \cite{Louko:2006zv,Satz:2006kb} (also see \cite{Louko:2007mu}) with special care for the modifications arising from the spatial compactification. With the mathematical tools at hand, the response of the Unruh-DeWitt detector is investigated in depth for three different settings: \secref{sec:constant velocity} for the detector moving at a constant velocity, \secref{sec:acceleration in compact direction} for moving with a constant acceleration in the compact direction, and \secref{sec:acceleration in noncompact directions} for moving with a constant acceleration in noncompact directions. Finally, the results and their implications are summarized and discussed in \secref{sec:discussion}.
Throughout this paper, we use the metric signature $(-,+,+,+)$ and the natural units with both $\hbar$ and the speed of light set to be 1. (The symbol $c$ is used to denote the coupling constant for the detector's interaction with the scalar field.)

\section{The Unruh-DeWitt detector}\label{sec:Unruh-DeWitt detector}
In the generally covariant description of quantum field theory, the notion of ``particles'' of the quantum field is rather ambiguous in the sense that the particle content is observer-dependent \cite{Fulling:1972md}. It is therefore natural to seek out an operational definition of particles in terms of the response of a well-defined ``particle detector''.
The model of photon detectors has been considered in the context of quantum optics by Glauber in 1963 \cite{Glauber:1963fi}. However, it was not until Unruh introduced a particle detector model in 1976 that the problem of a detector's response in relation to its trajectory was undertaken \cite{Unruh:1976db}. Unruh's model is given by a particle in a small box coupled to the quantum field and a particle of the quantum filed is said to be detected if the particle in the box is excited from its initial ground state to some excited state. (A similar model was also developed by S\'{a}nchez in 1981 \cite{Sanchez:1981xx}.) In 1979, DeWitt \cite{DeWitt:1979} further improved Unruh's idea by introducing a two-level point monopole detector, which is now generally referred to as the Unruh-DeWitt detector in the literature.

In the following, we briefly review the Unruh-DeWitt detector, particularly following the lines of Sec.\ 3.3 in \cite{Birrell:1982ix} and using the same notations as closely as possible. Unlike \cite{Birrell:1982ix}, we consider the transition rates of both excitation ($\Delta E>0$) and de-excitation ($\Delta E<0$). We also take into account the switching function in the form as introduced in \cite{Louko:2006zv,Satz:2006kb,Louko:2007mu}. More details about the Unruh effect and the Unruh-DeWitt detector can be found in \cite{Birrell:1982ix,Wald:book,Padmanabhan:2003gd} and especially \cite{Crispino:2007eb}.

The Unruh-DeWitt detector is an idealized point-particle detector with two energy levels $\ket{E_0}$ and $\ket{E}$, coupled to a scalar field $\phi$ via a monopole interaction. If the detector moves along a world line $x^\mu(\tau)$, where $\tau$ is the detector's proper time, the Lagrangian for the monopole interaction is given by
\begin{equation}\label{monopole  interaction}
c\,\chi(\tau)\mu(\tau)\phi(x^\mu(\tau)),
\end{equation}
where $c$ is a small coupling constant, $\mu(\tau)$ is the operator of the detector's monopole moment, and $\chi(\tau)$ is a switching function with a compact support (i.e., positive for a finite period of time and zero before and after the interaction), accounting for the switch-on and switch-off of the interaction.

For a generic trajectory $x^\mu(\tau)$, the detector in general does not remain in its initial state $\ket{E_0}$ but can be excited (if $\Delta E := E-E_0 >0$) or de-excited (if $\Delta E<0$) to the other state $\ket{E}$, while at the same time the field $\phi$ makes a transition from the vacuum state $\ket{0}$ to an excited state $\ket{\Psi}$.\footnote{See \appref{app:equilibrium} for the confusion arising from the different usages of the words \emph{excitation} and \emph{de-excitation}.} By the first-order perturbation theory, the amplitude for the transition $\ket{0,E_0}\rightarrow\ket{\Psi,E}$ is given by
\begin{equation}\label{transition amplitude 0}
ic\,\bra{\Psi,E}\int_{-\infty}^\infty \chi(\tau) \mu(\tau)\, \phi\left(x^\mu(\tau)\right) d\tau \ket{0,E_0},
\end{equation}
which leads to the factorized form:
\begin{equation}\label{transition amplitude}
ic \bra{E}\mu(0)\ket{E_0} \int_{-\infty}^\infty e^{i(E-E_0)\tau} \chi(\tau) \bra{\Psi}\phi\left(x^\mu(\tau)\right)\ket{0}\,d\tau
\end{equation}
by the equation of evolution for $\mu(\tau)$:
\begin{equation}
\mu(\tau)=e^{iH_0\tau}\mu(0)e^{-iH_0\tau}.
\end{equation}
Summing the squared norm of the amplitude given in \eqnref{transition amplitude} over all possible $\ket{\Psi}$,\footnote{Here, we use the completeness relation $\sum_{\ket{\Psi}}\ket{\Psi}\bra{\Psi}=\mathbbm{1}$, but note that, at the level of the first-order perturbation, only the one-particle states of $\ket{\Psi}$ contribute.} we obtain the transition probability of $\ket{E_0}\rightarrow\ket{E}$ as
\begin{equation}\label{transition probability}
c^2\abs{\bra{E}\mu(0)\ket{E_0}}^2\ F(E-E_0),
\end{equation}
where
\begin{equation}\label{response function}
F(\Delta E) = \int_{-\infty}^\infty d\tau \int_{-\infty}^\infty d\tau'
e^{-i\Delta E(\tau-\tau')} \chi(\tau)\,\chi(\tau')\, G^+(x(\tau),x(\tau'))
\end{equation}
is the \emph{response function}, which depends on the trajectory but not the internal properties of the detector. The remaining factor $c^2\abs{\bra{E}\mu(0)\ket{E_0}}^2$ represents the \emph{selectivity}, which depends only on the detector's internal properties.\footnote{In the rest of the paper, we will focus on the response function $F(\Delta E)$ and ignore the factor of selectivity.} The Wightman functions $G^{\pm}$ are defined as
\begin{subequations}
\begin{eqnarray}
G^+(x,x')&:=&\bra{0}\phi(x)\phi(x')\ket{0},\\
G^-(x,x')&:=&\bra{0}\phi(x')\phi(x)\ket{0}.
\end{eqnarray}
\end{subequations}

The detector is said to be in equilibrium with the field $\phi$ along a given trajectory, if
\begin{equation}\label{equilibrium}
G^+(\tau,\tau')\equiv G^+(x(\tau),x(\tau')) = G^+(\Delta\tau), \quad \Delta\tau:=\tau-\tau',
\end{equation}
depends only on $\Delta\tau$.
In this case, imposition of the switching function $\chi(\tau)$ can be viewed as a prescription of regularization. Correspondingly, by trivially prescribing $\chi(\tau)=1$, the (infinite) total transition probability divided by the (infinite) total proper time gives rise to the (finite) \emph{equilibrium transition rate} (i.e., probability per unit proper time) given by
\begin{equation}\label{transition rate}
R=
c^2\abs{\bra{E}m(0)\ket{E_0}}^2
\dot{F}(\Delta E),
\end{equation}
where
\begin{equation}\label{dot F}
\dot{F}(\Delta E) := \int_{-\infty}^\infty d(\Delta\tau)
e^{-i\Delta E\Delta\tau} G^+(\Delta\tau).
\end{equation}

On the other hand, if the detector is not in equilibrium with $\phi$ (i.e., $G^+(\tau,\tau')$ depends on both $\tau$ and $\tau'$ for the given trajectory), we are unable to make sense of the notion of equilibrium transition rate but can only refer to the \emph{total} transition probability, which now depends on the exact form of $\chi(\tau)$. Provided that $\chi(\tau)$ is smoothly switched on for a finite duration of time and then smoothly switched off and that the coupling constant $c$ is small enough in comparison to the switch-on duration (so that the first-order perturbation is viable), \eqnref{transition probability} with \eqnref{response function} is well defined and yields a finite total transition probability. By taking the time derivative of the total transition probability, we can still define the \emph{instantaneous transition rate} observed at a particular instant, the notion of which will be elaborated in \secref{sec:sharp switching limit}.

Also see \appref{app:equilibrium} for the remarks on how the transition rate can be measured in principle and what the condition of equilibrium is and is not.

\section{The Wightman function}\label{sec:Wightman function}
Consider a real scalar field $\phi(x)\equiv\phi(t,\mathbf{x})$ in a $d$-dimensional spacetime, where events are coordinated as $x^\mu=(x^0,\mathbf{x})=(t,x^1,\dots,x^{d-1})$. The mode expansion of $\phi(x)$ is given by
\begin{equation}
\phi(t,\mathbf{x})=\sum_\mathbf{k}
\left(a_\mathbf{k}u_\mathbf{k}(t,\mathbf{x})
+a_\mathbf{k}^\dag u_\mathbf{k}^*(t,\mathbf{x})\right).
\end{equation}
If the spacetime is flat but the $(d-1)$-th spatial direction is compactified with a finite length $L$, the Fourier modes $u_\mathbf{k}$ are given by
\begin{equation}
u_\mathbf{k}(t,\mathbf{x}) = \frac{1}{\left(2\omega_\mathbf{k}(2\pi)^{d-2}L\right)^{1/2}}\,
e^{i\mathbf{k}\cdot\mathbf{x}-i\omega_\mathbf{k}t},
\end{equation}
where the frequency associated with $\mathbf{k}=(k^1,\dots,k^{d-1})$ is
\begin{equation}
\omega_\mathbf{k} := \sqrt{\mathbf{k}^2+m^2},
\end{equation}
and the $(d-1)$-th component of $\mathbf{k}$ takes only discrete values:
\begin{equation}
k^{d-1}=\frac{2\pi n}{L}, \quad n\in\mathbb{Z}.
\end{equation}

Let $\ket{0_L}$ be the vacuum state in accordance with the above mode expansion, i.e.,
\begin{equation}
a_\mathbf{k}\ket{0_L} = 0, \quad \text{for}\ \forall\,\mathbf{k}.
\end{equation}
The Wightman function $G_L^+(x,x')$ then takes the form
\begin{eqnarray}\label{GL+}
G_L^+(x,x') &:=& \bra{0_L}\phi(x)\phi(x')\ket{0_L} \nonumber\\
&=& \left(\frac{1}{L}\sum_{k^{d-1}\in\frac{2\pi}{L}\mathbb{Z}}\right)
\int \frac{d^{d-2}\mathbf{k}}{(2\pi)^{d-2}}
\frac{1}{2\omega_\mathbf{k}}\,
e^{i\mathbf{k}\cdot(\mathbf{x}-\mathbf{x'})-i\omega_\mathbf{k}(t-t')} \nonumber\\
&=& \sum_{n=-\infty}^{\infty}
\int \frac{d^{d-1}\mathbf{k}}{(2\pi)^{d-1}}
\frac{1}{2\omega_\mathbf{k}}\,
e^{i\mathbf{k}\cdot(\mathbf{x}-\mathbf{x'})-i\omega_\mathbf{k}(t-t')}
e^{-inLk^{d-1}},
\end{eqnarray}
where we have used the Poisson summation formula.\footnote{The Poisson summation formula is \begin{equation*}
\sum_{n=-\infty}^\infty f(n)= \sum_{k=-\infty}^\infty \int_{x=-\infty}^\infty dx\, f(x) e^{-2\pi i k x}.
\end{equation*}} The Wightman function depends only on the difference of $x$ and $x'$, i.e., $G_L^+(x,x')=G_L^+(x-x')$; furthermore, as can be seen from \eqnref{GL+}, it is periodic in the $x^{d-1}$ direction, i.e.,
\begin{equation}
G_L^+(t'-t',x^1-x^{1'},\dots,x^{d-1}-x^{d-1'}+nL) = G_L^+(t'-t',x^1-x^{1'},\dots,x^{d-1}-x^{d-1'}),
\quad n\in\mathbb{Z}.
\end{equation}
Eq.~\eqnref{GL+} can be cast as
\begin{equation}\label{GL+ 2}
G_L^+(x,x') = \sum_{n=-\infty}^\infty G^+(t-t',x^1-x^{1'},\dots,x^{d-1}-x^{d-1'}-nL),
\end{equation}
where $G^+(x,x')\equiv G_{L\rightarrow\infty}^+(x,x')$ is the ordinary Wightman function in the Minkowski spacetime. Eq.~\eqnref{GL+ 2} is a known result and can also be obtained by the method of images (e.g., see \cite{Davies:1989me}).

The Green functions (Wightman function included) are generally very complicated. In the case of a massless ($m=0$) scalar field in 4-dimensional spacetime, $G^+(x,x')$ can be explicitly calculated as (see \cite{Birrell:1982ix,Wald:book,Padmanabhan:2003gd,Crispino:2007eb})
\begin{equation}\label{G+}
G^+(x,x') = -\frac{1}{4\pi^2}\, \frac{1}{(t-t'-i\epsilon)^2-\abs{\mathbf{x}-\mathbf{x'}}^2},
\end{equation}
where we adopt the standard $i\epsilon$-regularization by introducing a small (infinitesimal) imaginary number $i\epsilon$, $\epsilon>0$, to provide a frequency cutoff.
The rest of this paper will focus on this 4-dimensional case with $x^\mu=(t,x,y,z)$.

Once the detector's trajectory is given, the Wightman function is known and we are ready to compute the transition rate for the equilibrium case by \eqnref{dot F} with $G^+(\Delta \tau)$ replaced by $G_L^+(\Delta \tau)$. For the nonequilibrium case, however, we have to take more care with the switching function in order to make sense of the instantaneous transition rate.

\section{Transition probability with a switching function}\label{sec:with switching}
The response function given by \eqnref{response function} involves the regularization $i\epsilon$, which is prescribed in the Wightman function $G^+(x,x')$ as shown in \eqnref{GL+ 2} and \eqnref{G+}. Provided that the trajectory $x(\tau)$ is smooth enough,\footnote{See \footref{footnote:C9} for the precise requirement.} \eqnref{response function} can be recast in a form free of regularization, which can then be computed explicitly (at least numerically). This can be done by following the same steps in \cite{Louko:2006zv,Satz:2006kb} (also see \cite{Louko:2007mu}) except that we should take special care of the modifications arising from the spatial compactification.

The detailed derivation is given in \appref{app:detail 1}. The regularization-free expression of \eqnref{response function} turns out to be
\begin{eqnarray}\label{F final form}
F(\Delta E)
&=&
-\frac{\Delta E}{4\pi} \int_{-\infty}^\infty du\, \chi(u)^2 \nonumber\\
&& \mbox{}
+ \frac{1}{2\pi^2}\int_{-\infty}^\infty du\,\chi(u)\int_0^\infty ds\,
\chi(u-s)\left(\cos(\Delta E s)\sum_{n=-\infty}^\infty\frac{1}{\Delta x_n^2}
+\frac{1}{s^2}\right) \nonumber\\
&& \mbox{}
+\frac{1}{2\pi^2} \int_0^\infty \frac{ds}{s^2}
\int_{-\infty}^\infty du\, \chi(u)\left(\chi(u)-\chi(u-s)\right),
\end{eqnarray}
where we define
\begin{subequations}\label{Dt and Dxn}
\begin{eqnarray}
\Delta t &:=& t(u)-t(u-s)\\
\Delta\mathbf{x}_n^2 &:=&
\big(x(u)-x(u-s)\big)^2 + \big(y(u)-y(u-s)\big)^2 + \big(z(u)-z(u-s)-nL\big)^2.
\end{eqnarray}
\end{subequations}
and
\begin{subequations}\label{Dx and Dxn}
\begin{eqnarray}
\label{Dxn square}
\Delta x_n^2 &:=& -(\Delta t)^2 + \Delta\mathbf{x}_n^2
= (\Delta x)^2 -2nL\Delta z +(nL)^2,\\
\label{Dx square}
(\Delta x)^2 &:=& \Delta x^\mu \Delta x_\mu
\equiv\big(x^\mu(u)-x^\mu(u-s)\big)\big(x_\mu(u)-x_\mu(u-s)\big) \nonumber \\
&\equiv& -(\Delta t)^2+(\Delta x)^2+(\Delta y)^2+(\Delta z)^2.
\end{eqnarray}
\end{subequations}
The factor
\begin{equation}\label{sum factor}
\sum_{n=-\infty}^\infty \frac{1}{\Delta x_n^2(u,s)} \equiv
\sum_{n=-\infty}^\infty \frac{1}{(\Delta x)^2-2nL\Delta z+(nL)^2}
\equiv 4\pi^2 G_L^+(u,u-s)\Big|_{\epsilon=0}
\end{equation}
appearing in \eqnref{F final form} can be cast in a closed form via the identity\footnote{The right-hand side of \eqnref{identity a} always yields a real number even if $a^2-4b<0$, as in this case it is easy to show that both the numerator and denominator are purely imaginary.}
\begin{equation}\label{identity a}
f(a,b):=
\sum_{n=-\infty}^\infty \frac{1}{n^2+an+b} =
\frac{\pi \cot \left[\frac{\pi}{2}
   \left(a-\sqrt{a^2-4b}\right)\right]
   -\pi \cot
   \left[\frac{\pi}{2} \left(
   a+\sqrt{a^2-4 b}
   \right)\right]}{\sqrt{a^2-4 b}}.
\end{equation}
It should be emphasized that \eqnref{F final form} is valid only if the factor \eqnref{sum factor} has no singular points other than $s=0$ along the detector's trajectory. On the other hand, if the factor \eqnref{sum factor} has singular points other than $s=0$ along the given trajectory, the derivation in \appref{app:detail 1} breaks down and \eqnref{F final form} is viable only if the switching function $\chi(u-s)$ is turned off before \eqnref{sum factor} hits a singularity other than $s=0$.\footnote{In any case, one can always compute the transition probability by \eqnref{F1}, but \eqnref{F1} will not reduce to the same simple regularization-free form of \eqnref{F final form} unless $\chi(u-s)$ is turned off before \eqnref{sum factor} hits any nonzero singularities.}

Note that the improper integrals over $(0,\infty)$ and $(-\infty,\infty)$ appearing in \eqnref{F final form} are in fact only over finite intervals, since $\chi(u)$ is of compact support. Furthermore, since $\Delta x_{n=0}^2\equiv (\Delta x)^2 = -s^2+O(s^4)$ (see \appref{app:detail 1}), the divergent behavior of $1/s^2$ inside the parenthesis in \eqnref{F final form} as $s\rightarrow0$ is cancelled out by the same divergent behavior of the term $\cos(\Delta Es)/\Delta x_{n=0}^2$. Therefore, if the factor \eqnref{sum factor} has no singular points other than $s=0$ along the detector's trajectory, the expression \eqnref{F final form} is not only free of regularization but yields a finite result. As long as the detector's trajectory $x^\mu(\tau)$ and the switching function $\chi(u)$ are both explicitly given and smooth enough,\footnote{\label{footnote:C9}According to \cite{Satz:2006kb}, in order to obtain \eqnref{I< n=0} from \eqnref{I<}, the trajectory has to be of $C^9$ for smoothness.}$^,$\footnote{A function is said to be smooth enough, if it is of class $C^n$ to a certain degree $n>0$. One can choose $\chi(u)$ to be smooth (i.e., of $C^\infty$) by modeling it as a bump function. However, $\chi(u)$ cannot be analytic, since a nonzero analytic function cannot be compactly supported.} and furthermore the factor \eqnref{sum factor} is free of singularities other than $s=0$ along the trajectory, \eqnref{F final form} gives an unambiguous finite result for the total transition probability.

In the limit $L\rightarrow\infty$, all the summands in \eqnref{sum factor} vanish except for the $n=0$ term. Consequently, $\sum_{n=-\infty}^\infty 1/\Delta x_n^2 \rightarrow 1/(\Delta x)^2$ and \eqnref{F final form} reduces to the ordinary result of the Minkowski spacetime as given in (3.8) of \cite{Satz:2006kb}. Also note that in this case, the factor \eqnref{sum factor} is always free of singular points except for $s=0$ regardless of the detector's trajectory.

\section{The sharp switching limit}\label{sec:sharp switching limit}
\begin{figure}
\centering
    \includegraphics[]{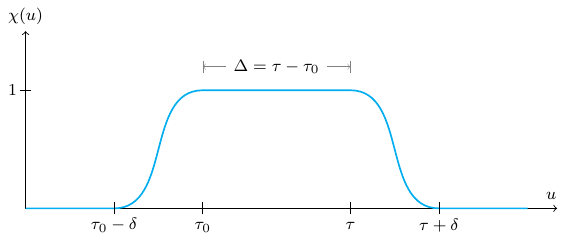}
\caption{A typical switching function $\chi(u)$.}
\label{fig:switching function}
\end{figure}

The switching function $\chi(u)$ can be modeled as a smooth enough function as shown in \figref{fig:switching function}.
That is, $\chi(u)$ is 0 for $u<\tau_0-\delta$; it is smoothly turned on from 0 to 1 during $u\in(\tau_0-\delta,\tau_0)$; it remains to be 1 for $u\in(\tau_0,\tau)$; it is then smoothly turned off from 1 to 0 during $u\in(\tau,\tau+\delta)$; and finally it remains to be 0 for $u>\tau+\delta$. The interval $\Delta:=\tau-\tau_0$ is understood as the switch-on duration and $\delta$ indicates how fast the switch-on and -off are performed. To neglect the detailed dependence on the switching function, it is convenient to take the artificial prescription of the sharp switching limit given by $\delta/\Delta\rightarrow0$.\footnote{In the sharp switching limit, however, $\chi(u)$ is no longer smooth enough, which gives rise to an artifact of logarithmic divergence in the transition probability as we will see shortly.} We follow the same steps in \cite{Satz:2006kb} to obtain the sharp switching limit of \eqnref{F final form}.

In the sharp switching limit $\delta/\Delta\rightarrow0$, $\chi(u)$ becomes $\Theta(u-\tau_0)\Theta(u-\tau)$ with $\Theta(x)$ being the Heaviside step function. The first term in \eqnref{F final form} reduces to
\begin{equation}
-\frac{\Delta E}{4\pi} \int_{-\infty}^\infty du\, \chi(u)^2 = -\frac{\Delta E}{4\pi} \Delta
+ O\left(\frac{\delta}{\Delta}\right).
\end{equation}
The second term reduces to
\begin{eqnarray}
&& \frac{1}{2\pi^2}\int_{-\infty}^\infty du\,\chi(u)\int_0^\infty ds\,
\chi(u-s)\left(\cos(\Delta E s)\sum_{n=-\infty}^\infty\frac{1}{\Delta x_n^2}
+\frac{1}{s^2}\right) \nonumber\\
&=& \frac{1}{2\pi^2}\int_{\tau_0}^\tau du \int_0^{u-\tau_0} ds\,
\left(\cos(\Delta E s)\sum_{n=-\infty}^\infty\frac{1}{\Delta x_n^2}
+\frac{1}{s^2}\right)
+ O\left(\frac{\delta}{\Delta}\right).
\end{eqnarray}
The third term however yields
\begin{equation}
\frac{1}{2\pi^2} \int_0^\infty \frac{ds}{s^2}
\int_{-\infty}^\infty du\, \chi(u)\left(\chi(u)-\chi(u-s)\right)
= \frac{1}{2\pi^2} \int_0^\Delta \frac{ds}{x} + O\left(\frac{\delta}{\Delta}\right),
\end{equation}
which is logarithmically divergent. The logarithmic divergence is in fact an artifact due to the infinite sharpness of the switching. By the detailed calculation in Sec.~4 of \cite{Satz:2006kb}, the logarithmic divergence can be rendered more explicitly and in the end the response function takes the form:
\begin{eqnarray}\label{F sharp}
F(\Delta E) &=&
-\frac{\Delta E}{4\pi}(\tau-\tau_0)
+\frac{1}{2\pi^2}\int_{\tau_0}^\tau du \int_0^{u-\tau_0} ds\,
\left(\cos(\Delta E s)\sum_{n=-\infty}^\infty\frac{1}{\Delta x_n^2(u,s)}
+\frac{1}{s^2}\right) \nonumber\\
&& \mbox{} + \frac{1}{2\pi^2}\ln \frac{\Delta}{\delta}
+ C + O\left(\frac{\delta}{\Delta}\right),
\end{eqnarray}
where $C$ is a constant independent of $\Delta E$, $\Delta$, and $\delta$. Taking the derivative with respect to $\tau$ upon \eqnref{F sharp} yields
\begin{eqnarray}\label{dot F tau}
\dot{F}_\tau(\Delta E) := \frac{d}{d\tau} F(\Delta E)
&=& -\frac{\Delta E}{4\pi}
+\frac{1}{2\pi^2}\int_0^\Delta ds
\left(\cos(\Delta E s)\sum_{n=-\infty}^\infty\frac{1}{\Delta x_n^2(\tau,s)}
+\frac{1}{s^2}
\right) \nonumber\\
&& \mbox{} + \frac{1}{2\pi^2\Delta} + O\left(\frac{\delta}{\Delta^2}\right),
\end{eqnarray}
where $\sum_{n=-\infty}^\infty 1/\Delta x_n^2(\tau,s)$ is given by \eqnref{sum factor} with $u$ replaced by $\tau$. Eq.~\eqnref{dot F tau} (together with the factor of selectivity) can be understood as the \emph{instantaneous transition rate} as observed at the instant $\tau$ for the detector being sharply turned on at the instant $\tau_0$. Even though the total transition probability given by $F(\Delta E)$ is divergent and thus ill defined in the sharp switching limit, the instantaneous transition rate in the limit $\delta\rightarrow 0$ remains well-behaved everywhere except for $\tau=\tau_0$ (see \footref{foot:pathological}).

In the case that \eqnref{sum factor} has singularities other than $s=0$, \eqnref{dot F tau} remains viable as long as the switch-on duration $\Delta=\tau-\tau_0$ is short enough such that $s\in(0,\Delta)$ encounters no nonzero singularities (that is, the switching function $\chi(u-s)$ is turned off before $s$ hits any nonzero singularities as discussed in \secref{sec:with switching}). When $\Delta$ keeps increasing, the instantaneous transition rate $\dot{F}_\tau(\Delta E)$ might become divergent at a certain critical point at which $s$ hits a nonzero singularity. (We will see this case in more detail in \secref{sec:acceleration in compact direction}.) Finally, when $\Delta$ is larger than the critical value, \eqnref{dot F tau} becomes invalid.\footnote{This does not mean we can no longer make sense of the instantaneous transition rate, but just means \eqnref{F final form} is not viable and thus neither is \eqnref{dot F tau}. To derive the valid counterparts of \eqnref{F final form} and \eqnref{dot F tau}, we have to redo the analysis in \appref{app:detail 1} by taking the small $s$ expansion of \eqnref{sum factor} around the nonzero singularities, which is much more involved and depends on the explicit form of \eqnref{sum factor}.}

The formula \eqnref{dot F tau} is explicitly \emph{causal} in the sense that the instantaneous transition rate observed at time $\tau$ depends only on the detector's trajectory \emph{before} $\tau$ and independent of the trajectory \emph{after} $\tau$. (See \cite{Louko:2006zv,Schlicht:2003iy} for more discussions on the causality.) Although we have to assume a switch-off to obtain \eqnref{dot F tau}, the causality implies that the same formula of \eqnref{dot F tau} gives the instantaneous transition rate observed at $\tau$ regardless of whether the detector is turned off or remains turned on after the observation is made. If the detector follows a trajectory in equilibrium with the field $\phi$ and \eqnref{sum factor} is free of singularities other than $s=0$, we expect that, in the $\Delta\rightarrow\infty$ limit (i.e., the detector has been turned on for a long time), $\dot{F}_\tau(\Delta E)$ given in \eqnref{dot F tau} will asymptote to a $\tau$-independent value, which should be identical to $\dot{F}(\Delta E)$ given in \eqnref{dot F}. That is, the instantaneous transition rate becomes the equilibrium transition rate if the detector is turned on long enough. If \eqnref{sum factor} suffers from singularities other than $s=0$, it should be caveated that \eqnref{dot F tau} is valid only for short $\Delta$ and the $\Delta\rightarrow\infty$ limit of \eqnref{dot F tau} does not make sense; the equilibrium transition rate should be calculated directly by \eqnref{dot F}.

On the other hand, if the detector's trajectory is not in equilibrium with $\phi$, the instantaneous transition rate $\dot{F}_\tau(\Delta E)$ does not converge to an asymptotic value as $\Delta\rightarrow\infty$ (even if we can derive the valid counterpart of \eqnref{dot F tau} for large $\Delta$). In this case, instead of the equilibrium transition rate, we can only make sense of the instantaneous transition rate, which depends on the observation time $\tau$ and how long the detector has been turned on. The instantaneous transition rate for short $\Delta$ is given by \eqnref{dot F tau}.

It should be noted that the instantaneous transition rate might be negative for some values of $\tau$. This is not a pathological trait, because what is required to be positive is the \emph{total} transition probability, not its time derivative, i.e., the instantaneous transition rate. The fact that the instantaneous transition rate becomes negative around some instant $\tau$ simply means that the total transition probability becomes smaller if one chooses not to measure the signal at the instant $\tau$ but to wait and keep the detector turned on for a bit longer until the measurement is made at the later instant $\tau+d\tau$. It is the quantum interference over time that is responsible for the possibility that the total transition probability might not be monotonic against the switch-on duration.
However, if the detector's trajectory is in equilibrium with $\phi$, we can make sense of the \emph{equilibrium} transition rate as the $\Delta\rightarrow\infty$ limit of the instantaneous transition rate, which must be positive.

Similarly, it is also not pathological if the instantaneous transition rate is divergent at some critical points of $\tau$ in the case that \eqnref{sum factor} suffers from singularities other than $s=0$. Since the total transition probability is always finite (and positive), the instantaneous transition rate as a function of $\tau$ might be divergent at some points but must remain integrable over $\tau$.\footnote{\label{foot:pathological}The instantaneous transition rate given by \eqnref{dot F tau} does exhibit one pathological trait: it is divergent at the switch-on instant $\tau=\tau_0$ (i.e., $\Delta=0$). This divergence is non-integrable and gives rise to the logarithmic divergence in \eqnref{F sharp}, which is an artifact due to the infinite sharpness of the switching as discussed earlier.}

Finally, in the $\delta\rightarrow0$ limit, \eqnref{dot F tau} reveals an interesting relation between the excitation and de-excitation rates:
\begin{equation}\label{dot F diff}
\dot{F}_\tau(-\abs{\Delta E}) - \dot{F}_\tau(\abs{\Delta E})
= \frac{\abs{\Delta E}}{2\pi}.
\end{equation}
If \eqnref{sum factor} is free of nonzero singularities, \eqnref{dot F diff} is always held and even the equilibrium transition rate satisfies this relation, which is not obvious at all from \eqnref{dot F}.
On the other hand, if \eqnref{sum factor} suffers from nonzero singularities, \eqnref{dot F diff} is held only for short $\Delta$ and is broken down when $\Delta$ is larger than a certain critical value. Particularly, the equilibrium transition rate might violate this relation.

As we have obtained the formulae for both the equilibrium and instantaneous transition rates, we are ready to study the Unruh-DeWitt detector's response in various situation in the following three sections.

\section{Constant velocity}\label{sec:constant velocity}
Consider that the detector moves in the compact ($z$) direction at a constant velocity $\mathbf{v}=v_z\hat{z}$. The trajectory is given by the world line:
\begin{equation}
t=u^0\tau,\quad x=y=\text{const},\quad z=u^z\tau,
\end{equation}
where the 4-velocity $u^\mu$ is given by\footnote{Note that $u^0>u^z$ and $(u^0)^2-(u^z)^2=1$.}
\begin{equation}\label{u mu}
u^\mu = (u^0,u^x,u^y,u^z) = \left(\frac{1}{\sqrt{1-v_z^2}},0,0,\frac{v_z}{\sqrt{1-v_z^2}}\right).
\end{equation}
Eq.~\eqnref{GL+ 2} with \eqnref{G+} now reads as
\begin{eqnarray}\label{GL+ const v}
&&G_L^+(\tau,\tau') \equiv G_L^+(\Delta\tau)
= -\frac{1}{4\pi^2}\sum_{n=-\infty}^\infty \frac{1}{(u^0\Delta\tau-i\epsilon)^2-(u^z\Delta\tau-nL)^2} \nonumber\\
&=& -\frac{1}{4\pi^2}\sum_{n=-\infty}^\infty \frac{1}{\left((u^0+u^z)\Delta\tau-i\epsilon-nL\right) \left((u^0-u^z)\Delta\tau-i\epsilon+nL\right)}.
\end{eqnarray}
If $\Delta\tau$ is considered to be a complex number, each of the $n\neq0$ summands has two poles of order 1 at
\begin{equation}\label{simple poles}
\Delta\tau = \frac{nL+i\epsilon}{u^0+u^z},\quad
\frac{-nL+i\epsilon}{u^0-u^z},
\end{equation}
and for the $n=0$ summand the two simple poles merge into one pole of order 2 at
\begin{equation}\label{order-2 pole}
\Delta\tau = i\epsilon.
\end{equation}

For $\Delta E>0$, the equilibrium transition rate \eqnref{dot F} can be calculated by a contour integral along an infinite semicircle on the lower-half $\Delta\tau$ plane. However, as all poles in \eqnref{GL+ const v} are on the upper-half plane, the contour integral turns out to be zero.

For $\Delta E<0$, the equilibrium transition rate \eqnref{dot F} can be calculated by a contour integral along an infinite semicircle on the upper-half $\Delta\tau$ plane. The residue theorem gives
\begin{equation}\label{rate E<E0}
\dot{F}(\Delta E) =
-\frac{\Delta E}{2\pi}
-\frac{i}{4\pi L u^0} \sum_{n\in\mathbb{Z}-\{0\}}
\frac{1}{n}
\left(
e^{-i\Delta E \frac{nL}{u^0+u^z}}
-e^{i\Delta E \frac{nL}{u^0-u^z}}
\right),
\end{equation}
where the first term arises form the residue of the second-order pole in \eqnref{order-2 pole} and the remaining terms arise from the residues of the simple poles in \eqnref{simple poles}.
By the identity
\begin{equation}\label{identity b}
\sum_{n=1}^\infty \frac{e^{an}}{n} = - \ln (1-e^a),
\end{equation}
\eqnref{rate E<E0} can be cast in a closed form.

In summary, we have\footnote{Note that \eqnref{rate for constant velocity b} is real, as the numerators and denominators inside the parenthesis are complex conjugate to each other and thus the logarithmic function yields a purely imaginary number.}
\begin{subequations}\label{rate for constant velocity}
\begin{eqnarray}
\label{rate for constant velocity a}
\dot{F}(\Delta E) &=& 0, \qquad \text{for}\ \Delta E >0, \\
\label{rate for constant velocity b}
&=&
-\frac{\Delta E}{2\pi}
-\frac{i}{4\pi L u^0}
\ln
\left(
\frac{1-e^{i\frac{\Delta E L}{u^0+u^z}}}
{1-e^{-i\frac{\Delta E L}{u^0+u^z}}}
\,
\frac{1-e^{i\frac{\Delta E L}{u^0-u^z}}}
{1-e^{-i\frac{\Delta E L}{u^0-u^z}}}
\right),
\qquad \text{for}\ \Delta E <0.
\end{eqnarray}
\end{subequations}
In the limit $L\gg1/\Delta E$, as expected, \eqnref{rate for constant velocity} reduces to the ordinary result of the Minkowski spacetime:
\begin{equation}
\dot{F}(\Delta E)  \mathop{\longrightarrow}\limits_{L\rightarrow\infty}
-\frac{\Delta E}{2\pi}\Theta(-\Delta E).
\end{equation}
Note that this satisfies the relation \eqnref{dot F diff}.
%
%
For a generic value of $L$, the transition rate of excitation ($\Delta E>0)$ remains zero, while the transition rate of de-excitation ($\Delta E <0$) exhibits an extra correction, which depends on both $L$ and $u^z$, in addition to the ordinary result. Obviously, if we flip $u^z$ to $-u^z$ in \eqnref{rate for constant velocity b}, the transition rate remains the same, as it depends only on the magnitude of $u^z$, not the sign. It should also be noted that, even when $u^z=0$ (and $u^0=1$ correspondingly), the correction term in \eqnref{rate for constant velocity b} is nonzero.
Also note that, for a generic value of $L$, \eqnref{rate for constant velocity} violates the relation \eqnref{dot F diff}, essentially because \eqnref{sum factor} with \eqnref{GL+ const v} has singularities other than $s=0$.

Furthermore, although $\ket{0_L}$ is not invariant under the boost in the $z$ direction, it remains invariant under boosts in $x$ and $y$ directions. Therefore, the response of the detector should be the same if we boost it in $x$ and $y$ directions. That is, the response is still given by \eqnref{rate for constant velocity} with $u^0$ and $u^z$ still given by \eqnref{u mu}, even if the detector's velocity is $\mathbf{v}=v_x\hat{x}+v_y\hat{y}+v_z\hat{z}$ with $v_x,v_y\neq0$ in general.\footnote{The result of \eqnref{rate for constant velocity} for the case $u^z=0$ is the same as that of (4.2) in \cite{Davies:1989me} (with the replacements: $L\rightarrow2L$ for the different convention of $L$ and $\alpha\rightarrow0$ for the periodic boundary) computed by a different method.} The fact that the equilibrium transition rate of de-excitation depends on both $L$ and $u^z$ suggests that one can infer the $z$-component of the frame's moving velocity as well as the size of the compact direction by measuring $\dot{F}(\Delta E)$ of an Unruh-DeWitt detector for various values of $\Delta E<0$. That is, the nonequivalence of inertial frames is discernable by the response of an inertial Unruh-DeWitt detector.

If we are able to switch the detector on and off at will and measure the \emph{instantaneous} transition rate accordingly, then \emph{both} the excitation and de-excitation rates are nonzero and dependent on $L$ and $u^z$ (before the former asymptotes to zero). Thus, the instantaneous transition rates of both excitation and de-excitation can be used to deduce the $z$-component of the frame's velocity. This can be seen from the fact that the instantaneous transition rate of de-excitation must depend on $L$ and $u^z$ (otherwise its asymptotic value would not) plus the fact that the relation \eqnref{dot F diff} is still held for a short period of time. We can use \eqnref{dot F tau} to numerically compute the instantaneous transition rates of both kinds for a shot period of time, but the equilibrium transition rates as given in \eqnref{rate for constant velocity} cannot be obtained from \eqnref{dot F tau}.

In the next two sections, we will study the response of an accelerating detector.

\section{Constant acceleration in the compact direction}\label{sec:acceleration in compact direction}
Consider that the detector moves in the $z$ direction with a constant acceleration $1/\alpha$. The trajectory is given by the world line:
\begin{equation}\label{trajectory a in z}
t=\alpha\sinh\frac{\tau}{\alpha},\quad x=y=\text{const},\quad z=\alpha\cosh\frac{\tau}{\alpha}.
\end{equation}
Eq.~\eqnref{GL+ 2} with \eqnref{G+} now reads as
\begin{equation}\label{GL+ a in z original}
G_L^+(\tau,\tau')
=-\frac{1}{4\pi^2}\sum_{n=-\infty}^\infty \frac{1}{\left(\alpha\sinh\frac{\tau}{\alpha}-\alpha\sinh\frac{\tau'}{\alpha}-i\epsilon\right)^2
-\left(\alpha\cosh\frac{\tau}{\alpha}-\alpha\cosh\frac{\tau'}{\alpha}-nL\right)^2}.
\end{equation}
Consequently (see \appref{app:detail 2}), we have
\begin{equation}\label{GL+ a in z}
G_L^+(\tau,\tau')=-\frac{1}{16\pi^2\alpha^2}\sum_{n=-\infty}^\infty \frac{1}{\sinh^2\left(\frac{\Delta\tau}{2\alpha}-\frac{i\epsilon}{2\alpha}\right)
+n\left(\frac{L}{\alpha}\right)\sinh\frac{\tau+\tau'}{2\alpha}\sinh\frac{\Delta\tau}{2\alpha} -\frac{n^2}{4}\left(\frac{L}{\alpha}\right)^2}.
\end{equation}

When $L\gg \alpha$, only the summand with $n=0$  survives and \eqnref{GL+ a in z} reduces to the ordinary result of the Minkowski spacetime:
\begin{equation}\label{G+ a in z}
G^+(\tau,\tau')=-\frac{1}{16\pi^2\alpha^2} \frac{1}{\sinh^2\left(\frac{\Delta\tau}{2\alpha}-\frac{i\epsilon}{2\alpha}\right)}
=-\frac{1}{4\pi^2}\sum_{k=-\infty}^\infty \frac{1}{\left(\Delta\tau-i\epsilon+2\pi i\alpha k\right)^2},
\end{equation}
where we have used the identity
\begin{equation}\label{identity c}
\csc^2\pi x = \frac{1}{\pi^2}\sum_{k=-\infty}^\infty \frac{1}{(x-k)^2}.
\end{equation}
Taking \eqnref{G+ a in z} into \eqnref{dot F} and performing the contour integral, we obtain the transition rate given by
\begin{equation}\label{dot F thermal}
\dot{F}(\Delta E)
=\frac{\Delta E}{2\pi} \frac{1}{e^{2\pi\Delta E\alpha}-1}
\end{equation}
for both $\Delta E>0$ and $\Delta E<0$.
The celebrated Planck factor $(e^{2\pi\Delta E\alpha}-1)^{-1}$ indicates that the accelerated detector registers particles of $\phi$ as if it was immersed in a bath of thermal radiation at the temperature\footnote{However, see \cite{Crispino:2007eb}, especially Sec.\ III.A.4 therein, for more discussions about what the form of thermal radiation means and what it does not.}
\begin{equation}\label{temperature}
T=\frac{1}{2\pi k_\mathrm{B}\alpha}
\equiv \frac{\abs{\text{acceleration}}}{2\pi k_\mathrm{B}}.
\end{equation}
Also note that \eqnref{dot F thermal} satisfies the relation \eqnref{dot F diff}.

\begin{figure}

\centering
  \begin{minipage}[b]{0.48\textwidth}
    \includegraphics[width=\textwidth]{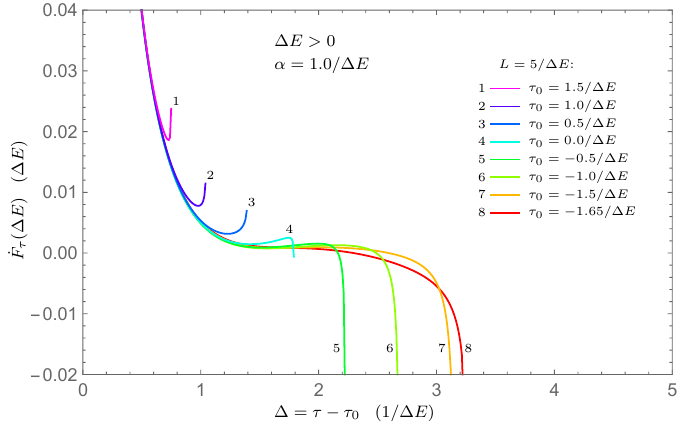}
  \end{minipage}
  \hspace{0.1cm} 
  \begin{minipage}[b]{0.48\textwidth}
    \includegraphics[width=\textwidth]{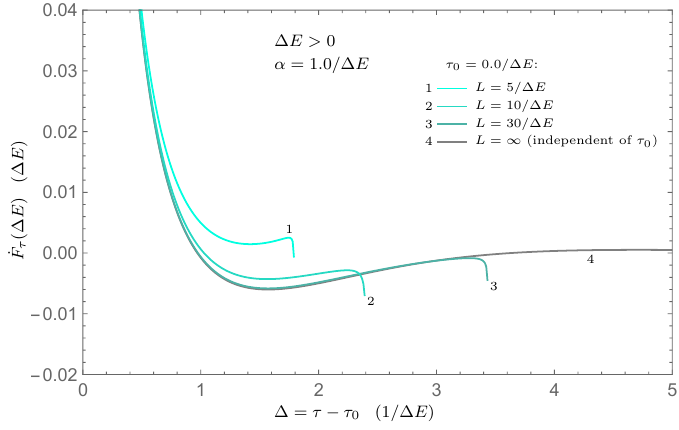}
  \end{minipage}

\caption{The instantaneous transition rate $\dot{F}_\tau(\Delta E)$ as a function of $\Delta$ for the trajectory \eqnref{trajectory a in z} with $\alpha=1.0/\Delta E$, $\Delta E>0$. \textbf{Left}: Curves for $L=5/\Delta E$ with various values of $\tau_0$. \textbf{Right}: Curves for $L=10/\Delta E$, $L=30/\Delta E$, and $L=\infty$ (uncompactified spacetime) with the same $\tau_0=0$. While the case of $L=\infty$ asymptotes to an equilibrium value, the other cases become divergent at critical points.}
\label{fig:transition rate}
\end{figure}

For generic cases that $L\not\gg \alpha$, \eqnref{GL+ a in z} can be cast in a closed form by the identity \eqnref{identity a}.
The closed form shows that $G_L^+(\tau,\tau')$ depends not only on $\Delta\tau\equiv\tau-\tau'$ but on both $\tau$ and $\tau'$, indicating that the detector moving in the $z$ direction with a constant acceleration is \emph{not} in equilibrium with the field $\phi$.\footnote{It is crucial to know whether the dependence on $\tau'$ in \eqnref{GL+ a in z} is erased away under the summation over $n$. By the identity \eqnref{identity a}, it is rigorously ascertained that $G_L^+(\tau,\tau')$ is dependent on both $\tau$ and $\tau'$.} As a consequence, instead of the equilibrium transition rate given by \eqnref{dot F}, we can only make sense of the \emph{instantaneous} transition rate, which is given by \eqnref{dot F tau} if the switch-on duration $\Delta$ is short enough that \eqnref{sum factor} encounters no singularities other than $s=0$. Taking \eqnref{trajectory a in z} into \eqnref{sum factor}, we have
\begin{eqnarray}\label{sum factor for a in z}
\sum_{n=-\infty}^\infty \frac{1}{\Delta x_n^2(\tau,s)}
&=& 4\pi^2 G_L^+(\tau,\tau-s)\Big|_{\epsilon=0}
= \frac{1}{L^2} f(a,b), \nonumber\\
\text{with}\quad a &=& -4\left(\frac{\alpha}{L}\right)
\sinh\left(\frac{2\tau-s}{2\alpha}\right) \sinh\left(\frac{s}{2\alpha}\right), \nonumber\\
b &=& -4\left(\frac{\alpha}{L}\right)^2 \sinh^2\left(\frac{s}{2\alpha}\right),
\end{eqnarray}
where $G_L^+(\tau,\tau')$ is given by \eqnref{GL+ a in z} and $f(a,b)$ is defined in \eqnref{identity a}.
With \eqnref{sum factor for a in z}, we now have a closed form for the integrand in \eqnref{dot F tau}. We do not have an analytic form for the whole expression \eqnref{dot F tau}, but it can be computed numerically by numerical integration. The numerical results of $\dot{F}_\tau(\Delta E)$ (with $\delta\rightarrow0$) as functions of $\Delta\equiv\tau-\tau_0$ are depicted in \figref{fig:transition rate} for various given values of $L$ and $\tau_0$ (the instant of switch-on).\footnote{\figref{fig:transition rate} only shows the instantaneous transition rate of excitation ($\Delta E>0$). The figure for the rate of de-excitation ($\Delta E<0$) is given as the same curves in \figref{fig:transition rate} but displaced upwards by $\abs{\Delta E}/2\pi$ according to \eqnref{dot F diff}. However, if we can manage to compute the instantaneous transition rate beyond the critical point, \eqnref{dot F tau} and consequently \eqnref{dot F diff} might break down as discussed previously.}
As expected, we can see the tendency from the figure that, as $L$ increases, $\dot{F}_\tau(\Delta E)$ becomes closer and closer to the ordinary result of the Minkowski spacetime (i.e., the result of $L\rightarrow\infty$). The ordinary result asymptotes to the equilibrium transition rate given by \eqnref{dot F thermal} as $\Delta\rightarrow\infty$. We also see that $\dot{F}_\tau(\Delta E)$, as a function of $\Delta\equiv\tau-\tau_0$, depends on both $L$ and $\tau_0$. Furthermore, $\dot{F}_\tau(\Delta E)$ can be negative (with $\Delta E>0)$ even for the ordinary ($L\rightarrow\infty$) result, but the asymptotic value of the ordinary result must be positive.

It calls our attention that the integrand in \eqnref{dot F tau} given with \eqnref{sum factor for a in z} blows up whenever $\cot[\cdots]$ in \eqnref{identity a} blows up. This happens when
\begin{equation}\label{condition for singularity}
a\pm\sqrt{a^2-4b}=2n,
\qquad
n\in\mathbb{Z}.
\end{equation}
In other words, the factor \eqnref{sum factor} has more singularities other than $s=0$, except for the case of $L\rightarrow\infty$. As a consequence, when $\Delta\equiv\tau-\tau_0$ becomes large enough, the integrand in \eqnref{dot F tau} will hit a singularity other than $s=0$ and the instantaneous transition rate will become divergent as discussed in \secref{sec:sharp switching limit}. We can see this divergent behavior in \figref{fig:transition rate} when $\tau-\tau_0$ approaches a certain critical point, beyond which the formula of \eqnref{dot F tau} is no longer valid. The exact value of the critical point agrees with the condition \eqnref{condition for singularity} with $\tau=\tau_0+\Delta$, $s=\Delta$.

Because $\ket{0_L}$ is invariant under arbitrary spacetime translations as well as Lorentz boosts in $x,y$ directions, the fact that $\dot{F}_\tau(\Delta E)$ depends on $\tau_0$ can be better understood as that it depends on the $z$-component of the \emph{instantaneous} 4-velocity $u^z(\tau_0):=\left.dz(\tau)/d\tau\right|_{\tau=\tau_0}=\sinh(\tau_0/\alpha)$ at the instant when the detector is turned on.
That is, within an initial reference frame moving at a constat velocity, if one instantaneously turns on and accelerates the Unruh-DeWitt detector in the $z$ direction, the detector's response depends not only on the magnitude of acceleration but also on the $z$-component of the frame's moving velocity.

\section{Constant acceleration in noncompact directions}\label{sec:acceleration in noncompact directions}
Consider that the detector moves with a constant acceleration $1/\alpha$ in noncompact directions (say, $x$ direction). The trajectory is given by the world line:
\begin{equation}
t=\alpha\sinh\frac{\tau}{\alpha},\quad y=z=\text{const},\quad x=\alpha\cosh\frac{\tau}{\alpha}.
\end{equation}
Eq.~\eqnref{GL+ 2} with \eqnref{G+} now reads as
\begin{eqnarray}\label{GL+ a in x}
G_L^+(\tau,\tau')
&=&-\frac{1}{4\pi^2}\sum_{n=-\infty}^\infty \frac{1}{\left(\alpha\sinh\frac{\tau}{\alpha}-\alpha\sinh\frac{\tau'}{\alpha}-i\epsilon\right)^2
-\left(\alpha\cosh\frac{\tau}{\alpha}-\alpha\cosh\frac{\tau'}{\alpha}\right)^2-n^2L^2} \nonumber\\
&=&-\frac{1}{16\pi^2\alpha^2}\sum_{n=-\infty}^\infty \frac{1}{\sinh^2\left(\frac{\Delta\tau}{2\alpha}-\frac{i\epsilon}{2\alpha}\right)
-\frac{n^2}{4}\left(\frac{L}{\alpha}\right)^2},
\end{eqnarray}
where the similar trick as shown in \appref{app:detail 2} has been repeated. Using the identity
\begin{equation}
\sum_{n=-\infty}^\infty \frac{1}{n^2-a^2} = -\frac{\pi \cot(\pi a)}{a}
\end{equation}
as a special case of \eqnref{identity a}, we can rewrite \eqnref{GL+ a in x} as
\begin{equation}\label{GL+ a in x 1}
G_L^+(\Delta\tau)=
-\frac{1}{8\pi\alpha L}\frac{\cot\left(\frac{2\pi\alpha}{L}
\sinh\left(\frac{\Delta\tau}{2\alpha}-\frac{i\epsilon}{2\alpha}\right)\right)}
{\sinh\left(\frac{\Delta\tau}{2\alpha}-\frac{i\epsilon}{2\alpha}\right)}.
\end{equation}
Note that \eqnref{GL+ a in x 1} reduces to \eqnref{G+ a in z} when $L\gg\alpha$.

To calculate the transition rate \eqnref{dot F} by the contour integral, we first rewrite \eqnref{GL+ a in x} as
\begin{eqnarray}\label{GL+ a in x 2}
G_L^+(\tau,\tau')
&=& -\frac{1}{16\pi^2\alpha^2} \frac{1}{\sinh^2\left(\frac{\Delta\tau}{2\alpha}-\frac{i\epsilon}{2\alpha}\right)}\nonumber\\
&&\mbox{}-\frac{1}{8\pi^2\alpha^2}\sum_{n=1}^\infty \frac{1}{\left(\sinh\left(\frac{\Delta\tau}{2\alpha}-\frac{i\epsilon}{2\alpha}\right)
-\frac{nL}{2\alpha}\right)
\left(\sinh\left(\frac{\Delta\tau}{2\alpha}-\frac{i\epsilon}{2\alpha}\right)
+\frac{nL}{2\alpha}\right)}.
\end{eqnarray}
The first part of \eqnref{GL+ a in x 2} is exactly the same as \eqnref{G+ a in z} and gives the same contribution as in \eqnref{dot F thermal}. For the second part, each of the summands has two singularities at
\begin{equation}\label{singularities}
\Delta\tau = \tau_{n,\pm} := i\epsilon\pm2\alpha\sinh^{-1}\frac{nL}{2\alpha}.
\end{equation}

For $\Delta E>0$, the transition rate \eqnref{dot F} can be calculated by a contour integral along an infinite semicircle on the lower-half $\Delta\tau$ plane. As the singularities in \eqnref{singularities} are all on the upper-half plane, only the first part of \eqnref{GL+ a in x 2} gives rise to a nonzero contour integral and consequently the transition rate is given by \eqnref{dot F thermal}.

For $\Delta E<0$, the transition rate \eqnref{dot F} is calculated by a contour integral along an infinite semicircle on the upper-half $\Delta\tau$ plane, and all the residues associated with the singularities in \eqnref{singularities} contribute.
Let
\begin{equation}
f(\Delta \tau) :=  \frac{e^{-i\Delta E\Delta\tau}}
{\left(\sinh\left(\frac{\Delta\tau}{2\alpha}-\frac{i\epsilon}{2\alpha}\right)
-\frac{nL}{2\alpha}\right)
\left(\sinh\left(\frac{\Delta\tau}{2\alpha}-\frac{i\epsilon}{2\alpha}\right)
+\frac{nL}{2\alpha}\right)}
=:\frac{g(\Delta\tau)}{h(\Delta\tau)}.
\end{equation}
Both $g$ and $h$ are holomorphic functions in a neighborhood of $\tau_{n,\pm}$.
We have $h(\tau_{n,\pm})=0$ and
\begin{equation}
h'(\Delta\tau) =
\frac{1}{\alpha}
\cosh\left(\frac{\Delta\tau}{2\alpha}-\frac{i\epsilon}{2\alpha}\right)
\sinh\left(\frac{\Delta\tau}{2\alpha}-\frac{i\epsilon}{2\alpha}\right),
\end{equation}
which follows
\begin{equation}
h'(\tau_{n,\pm}) = \pm\frac{nL}{2\alpha^2}
\sqrt{1+\left(\frac{nL}{2\alpha}\right)^2}
\neq 0,
\quad n=1,2,\dots,
\end{equation}
where we have used the identity $\cosh\left(\sinh^{-1}z\right)=\sqrt{1+z^2}$. As $h(\tau_{n,\pm})=0$ and $h'(\tau_{n,\pm})\neq0$, the residue of $f$ associated with $\tau_{n,\pm}$ can be computed as
\begin{equation}
\mathrm{Res}(f,\tau_{n,\pm}) =\lim_{\Delta\tau\rightarrow\tau_{n,\pm}} (\Delta\tau-\tau_{n,\pm}) f(\Delta\tau)
= \frac{g(\tau_{n,\pm})}{h'(\tau_{n,\pm})}
= \frac{\pm2\alpha^2 \exp\left(\mp2\alpha i \Delta E\sinh^{-1}\frac{nL}{2\alpha}\right)}
{nL\sqrt{1+\left(\frac{nL}{2\alpha}\right)^2}},
\end{equation}
where we have taken the $\epsilon\rightarrow0$ limit.
It then follows from the residue theorem that \eqnref{dot F} calculated by the semicircle contour integral leads to
\begin{eqnarray}\label{rate for a in x}
\dot{F}(\Delta E) &=& \frac{\Delta E}{2\pi} \frac{1}{e^{2\pi\Delta E\alpha}-1}
- \frac{\Theta(-\Delta E)}{8\pi^2\alpha^2} \sum_{n=1}^\infty 2\pi i\left(\mathrm{Res}(f,\tau_{n,+}) + \mathrm{Res}(f,\tau_{n,-})\right) \nonumber\\
&=& \frac{\Delta E}{2\pi} \frac{1}{e^{2\pi\Delta E\alpha}-1}
- \Theta(-\Delta E) \sum_{n=1}^\infty
\frac{\sin\left(2\alpha\Delta E\sinh^{-1}\frac{nL}{2\alpha}\right)}
{n\pi L\sqrt{1+\left(\frac{nL}{2\alpha}\right)^2}} \nonumber\\
&=:& \frac{\Delta E}{2\pi} \frac{1}{e^{2\pi\Delta E\alpha}-1}
- \Theta(-\Delta E) F(\Delta E,\alpha,L),
\end{eqnarray}
where we have inserted the step function to include both cases of $\Delta E>0$ and $\Delta E<0$.\footnote{We do not have a closed-form expression for the infinite series, but the comparison test tells that the series converges absolutely since
\begin{equation*}
\sum_{n=1}^\infty
\abs{
\frac{\sin\left(2\alpha\Delta E\sinh^{-1}\frac{nL}{2\alpha}\right)}
{n\pi L\sqrt{1+\left(\frac{nL}{2\alpha}\right)^2}}}
\leq \frac{2\alpha}{\pi L^2}\sum_{n=1}^\infty \frac{1}{n^2}
=\frac{\pi\alpha}{3L^2},
\end{equation*}
and we denote the converged value as $F(\Delta E,\alpha,L)$}$^,$\footnote{The result of \eqnref{rate for a in x} is the same as that of (4.4) in \cite{Davies:1989me} (with the replacements: $a\rightarrow1/\alpha$ for the acceleration, $L\rightarrow2L$ for the different convention of $L$, and $\alpha\rightarrow0$ for the periodic boundary) computed by a different method, except that the factor $-\Theta(-\Delta E)$ is wrongly assumed to be a thermal factor $(e^{2\pi\Delta E\alpha}-1)^{-1}$ in \cite{Davies:1989me}.}

It is crucial to know whether the dependence on $L$ in \eqnref{rate for a in x} goes away under the summation. By checking the limiting values of $F(\Delta E,\alpha,L)$:
\begin{subequations}
\begin{eqnarray}
F(\Delta E,\alpha,L) &\mathop{\longrightarrow}\limits_{L\rightarrow\infty}& 0,\\
&\mathop{\longrightarrow}\limits_{L\rightarrow0}& \sum_{n=1}^\infty \frac{\Delta E}{\pi} = -\infty,
\end{eqnarray}
\end{subequations}
it is ascertained  that the transition rate indeed depends on $L$ and therefore one can, in principle, infer the length $L$ by measuring the transition rate in relation to $\Delta E$ and $1/\alpha$.
On the other hand, the fact that $G^+_L(\tau,\tau')$ depends only on $\Delta\tau:=\tau-\tau'$, but not $\tau'$, means that the response of the detector cannot know about the frame's moving velocity in $x$, $y$ directions.

Finally, as a consistency check, the zero-accelerating limit ($\alpha\rightarrow\infty$) of \eqnref{rate for a in x} gives
\begin{equation}
\dot{F}(\Delta E)
\mathop{\longrightarrow}\limits_{\alpha\rightarrow\infty}
\Theta(-\Delta E)
\left(-\frac{\Delta E}{2\pi}
-\sum_{n=1}^\infty \frac{\sin(nL\Delta E)}{n\pi L}\right),
\end{equation}
which is identical to \eqnref{rate for constant velocity} with $u^z=0$ and $u^0=1$ as can be seen from \eqnref{rate E<E0}.
Also note that, for the same reason as discussed for the case moving at a constant velocity, if we are able to switch the detector on and off at will and measure the instantaneous transition rate, both the excitation and de-excitation rates will exhibit dependence on $L$.

\section{Summary and discussion}\label{sec:discussion}
If the Unruh-DeWitt detector moves at a constant velocity, the equilibrium transition rate of excitation ($\Delta E>0$) remains zero as the ordinary result in the Minkowski spacetime, but the equilibrium transition rate of de-excitation ($\Delta E<0$) receives an extra correction, which depends on the moving velocity in the compact ($z$) direction $u^z$ as well as the size of the compact dimension $L$, in addition to the ordinary result (the extra correction is nonzero even when $u^z=0$). If one can turn the detector on and off at will and measure the instantaneous transition rate, the rates of both excitation and de-excitation are nonzero (before the rate of excitation asymptotes to zero) and dependent on $u^z$ and $L$. The response of the Unruh-DeWitt detector can be used to infer $u^z$ and $L$ and therefore to discriminate between inertial frames with different velocities in the compact direction.

If the detector moves with a constant acceleration in the compact ($z$) direction, the response is not in equilibrium with the field $\phi$ and we can only make sense of the instantaneous transition rate. The instantaneous transition rates of both excitation and de-excitation depend not only on the acceleration $1/\alpha$ and the size $L$ but also on the time $\tau$ when the observation is made and the detector's instantaneous velocity $u^z(\tau_0)$ at the moment $\tau_0$ when the detector is turned on.
The detailed analysis also reveals that when $\Delta=\tau-\tau_0$ is large enough and approaches a critical point, the instantaneous transition rate becomes divergent, essentially due to the fact that the Wightman function has singularities other than zero in contrast to that of the ordinary Minkowski spacetime.

If the detector moves with an constant acceleration in noncompact ($x,y$) directions, the equilibrium transition rate of excitation remains to be the celebrated form of thermal radiation as in the ordinary Minkowski spacetime, while the equilibrium transition rate of de-excitation exhibits an extra correction dependent on $L$ in addition to the thermal form. Meanwhile, the frame's moving velocity in $x$, $y$ directions remains unknown. If we are able to measure the instantaneous transition rate, both the excitation and de-excitation rates will depend on $L$.

Comparing the results of the three different cases raises an issue concerning the condition of equilibrium. In the ordinary Minkowski spacetime, the condition for a detector to be in equilibrium with the background field is that its trajectory has to be \emph{stationary}. A trajectory is said to be stationary if the derivative with respect to its proper time is a Killing vector. For example, if \eqnref{trajectory a in z} is viewed as the $\xi=0$ curve in the coordinates $(\tau,x,y,\xi)$, which are related to $(t,x,y,z)$ via
\begin{equation}\label{tau and xi}
t = \alpha\, e^{\xi/\alpha}\sinh\frac{\tau}{\alpha},
\quad
z = \alpha\, e^{\xi/\alpha}\cosh\frac{\tau}{\alpha},
\end{equation}
then the vector field $\partial/\partial\tau\equiv \alpha^{-1}\left(t\partial/\partial z +z\partial/\partial t\right)$ is a Killing vector, which renders the metric invariant as we have
\begin{equation}\label{tau and xi 2}
ds^2=-\left(dt^2-dz^2\right)+dx^2+dy^2=
-e^{2\xi/\alpha}\left(d\tau^2 -d\xi^2\right) + dx^2 +dy^2.
\end{equation}
The coordinates $(\tau,x,y,\xi)$, however, do not cover the whole Minkowski spacetime but only the left and right Rindler wedges, i.e., the regions with $z^2>t^2$. The particle field can be expanded in terms of either the Minkowski modes or of the Rindler modes. Finding the Bogoliubov coefficients between these two expansions, one can deduce the fact that the Minkowski vacuum state is a thermal state of the Rindler modes at the temperature \eqnref{temperature} from the viewpoint of the left or right Rindler wedge.\footnote{\label{foot:Unruh effect}More precisely, tracing out the left (right) Rindler modes upon the Minkowski vacuum state gives rise to a density matrix for the many-particle system of the right (left) Rindler modes corresponding to the temperature \eqnref{temperature}. See \cite{Crispino:2007eb}, especially Eq.\ (2.78) therein, for more details.} Therefore, the transition rate of the Unruh-DeWitt detector obtained from the formula \eqnref{dot F} as being accompanied by the emission of a Minkowski-mode particle into the Minkowski vacuum can be reproduced from the Rindler observer perspective as being accompanied by the absorption of a Rindler-mode particle from the thermal bath (see Sec.~III.A.2 of \cite{Crispino:2007eb} for more details). More examples for other kinds of stationary trajectories and their corresponding Unruh-DeWitt transition rates can be found in Sec.~5.2 of \cite{Louko:2006zv}.

Rigorously speaking, however, the notion of ``stationarity'' is not a local concept. It makes no sense to call a \emph{single} trajectory stationary unless we are actually referring to a continuous \emph{family} of trajectories around the single one, because a Killing vector is in fact a vector \emph{field}, which cannot be associated with an isolated trajectory. If we view \eqnref{tau and xi} as a family of trajectories (parametrized by different values of $\xi$), under the spatial compactification, these trajectories serve as stationary coordinates only for a local spacetime region, but they fail to do so when the timescale is large enough because any given trajectory will eventually intersect with some other trajectories in the family due to $z\equiv z+nL$, $n\in\mathbb{Z}$. Put differently, we cannot find a proper sub-spacetime akin to the Rindler wedges from the global perspective and consequently we do not have the corresponding Bogoliubov coefficients. It seems to be the breakdown of the global stationarity that causes the detector moving with a constant acceleration in the compact direction not to be in equilibrium with $\phi$.
On the other hand, the case of a constant velocity and the case of a constant acceleration in noncompact directions do not suffer from the breakdown of the global stationarity, and both cases are in equilibrium with $\phi$.

Furthermore, we have two curious observations on the peculiar difference between the transition rates of excitation and de-excitation.
First, the transition rates of excitation and de-excitation satisfy a simple relation given by \eqnref{dot F diff}, which is broken down only when the Wightman function suffers from singularities other than zero and the switch-on duration $\Delta$ is long enough.
Second, in both \eqnref{rate for constant velocity} and \eqnref{rate for a in x}, the correction arising from the spatial compactification is perceptible only in the de-excitation rate but completely absent in the excitation rate. Somehow, the equilibrium transition rate of excitation is insensitive to the large-scale structure of spacetime, in contrast to the equilibrium rate of de-excitation and the instantaneous (nonequilibrium) rates of both kinds.
It is unclear whether these two observations and the aforementioned relation between equilibrium and stationarity remain true for generic settings; there seem to be some intriguing features about the Unruh effect not fully understood yet.

It should be remarked that whether we can definitely assert that inertial frames are discriminable by ``local'' experiments is a matter of interpretation. After all, the vacuum state $\ket{0_L}$ is a global concept and an Unruh-DeWitt detector knows about $\ket{0_L}$ only if the walls of the moving inertial reference frame are transparent to the field $\phi$.\footnote{In the same regard, the experiment by measuring the deviation of the electrostatic field of a point charge (as studied in \cite{Bansal:2005ue}) is not to be viewed as completely local either, since the deviation relies on the fact that the electric field stretches out to the entire universe and the local electric field is deformed anyway if the charge source is screened by the frame walls.} Finally, it should be noted that measurement of the response of the Unruh-DeWitt detector is completely out of reach of current technology, let alone able to distinguish the difference. Nevertheless, it is conceptually important to understand the (non)equivalence of inertial frames and that of uniformly accelerated frames in light of the response of the Unruh-DeWitt detector.


\begin{acknowledgments}
The author would like to thank Tun-Hao Chan (NTNU), Chia-Hsun Chang (NTNU), and En Shih (NTU) for having stimulating discussions, which inspired this work, and Thomas Roman (CCSU) for bringing related references to his attention. The substantial revision to the previous manuscript began as a response to an anonymous reviewer's suggestion and owes much to the discussion with Shih-Yuin Lin (NCUE). Another anonymous reviewer's constructive criticism also significantly helped to improve the manuscript. This work was supported in part by the Ministry of Science and Technology, Taiwan under the Grants No.\ 101-2112-M-003-002-MY3, No.\ 105-2811-M-003-028, and No.\ 106-2112-M-110-010.
\end{acknowledgments}


\appendix

\section{Remarks on equilibrium}\label{app:equilibrium}
The process of the detector's registering a signal of excitation or de-excitation can be represented as
\begin{subequations}\label{register}
\begin{eqnarray}
\label{register a}
\text{excitation:}\quad &&  \ket{0,E_0} \longrightarrow \ket{\Psi,E>E_0}, \\
\label{register b}
\text{de-excitation:}\quad &&  \ket{0,E_0} \longrightarrow \ket{\Psi,E<E_0},
\end{eqnarray}
\end{subequations}
where $\ket{\Psi}$ is some one-particle state of $\phi$ and $\ket{0}$ is the vacuum state. To measure the transition probability or transition rate corresponding to either of \eqnref{register}, we have to prepare a large ensemble of identical detectors (with the same coupling constant and the same switching function). To begin with, all the detectors are set to be in the state $\ket{E_0}$. Furthermore, because we are not interested in the reverse process of \eqnref{register}, i.e.,
\begin{subequations}\label{reverse process}
\begin{eqnarray}
&&  \ket{\Psi,E>E_0} \longrightarrow \ket{0,E_0}, \\
&&  \ket{\Psi,E<E_0} \longrightarrow \ket{0,E_0},
\end{eqnarray}
\end{subequations}
we should devise a ``halting'' mechanism so that whenever a detector registers a signal, it is turned off immediately and at the same time the one-particle state $\ket{\Psi}$ is reverted back to $\ket{0}$ by removing the extra particle emitted by the detector. If the above prescription can be achieved, then we can measure $N(\tau)$ as how many detectors are halted at a given time $\tau$ and compute the ratio
\begin{equation}\label{P(tau)}
P(\tau) = \frac{N(\tau)}{N_0},
\end{equation}
where $N_0$ is the total number of the ensemble. If the switching function $\chi(\tau)$ is of compact support, $P(\tau\rightarrow\infty)$ will yield the transition probability that is to be compared with \eqnref{response function} (times the selectivity). On the other hand, if each detector remains turned on until it registers a signal (i.e, $\chi(\tau)=1$ for $\tau>\tau_0$), all detectors will register signals sooner or later and be halted eventually. At the time when the number of unhalted detectors is still large enough, we can deduce the time derivative $\dot{P}(\tau)\equiv\dot{N}(\tau)/N_0$, which is reckoned to be the instantaneous transition rate. When the number of unhalted detectors becomes small, however, $\dot{P}(\tau)$ is no longer a faithful measure of the transition rate.

One can choose to maintain a large number of unhalted detectors, if a ``reviving'' mechanism is also devised. That is, among the halted detectors, we can choose to revive some of them by turning on the detectors again and resetting them to the initial state $\ket{E_0}$. The \emph{reviving rate} is said to be $\dot{Q}(\tau)$ if during the period from $\tau$ to $\tau+d\tau$, the probability of halted detectors being revived is given by $\dot{Q}(\tau)d\tau$. By adjusting the reviving rate, we can keep both the numbers of halted and unhalted detectors large enough. The instantaneous transition rate is then given by $\dot{N}(\tau)/N_0+\dot{Q}(\tau)$, which is to be compared with \eqnref{dot F tau} (times the selectivity). In the case that the trajectory $x^\mu(\tau)$ is in equilibrium with the background field, i.e., \eqnref{equilibrium} is satisfied, the reviving rate can be fine-tuned to match the equilibrium transition rate, which is manifested as the halting rate, such that $N(\tau)/N_0$ remains almost constant (up to small probabilistic fluctuations).  In the sense that \emph{the reviving rate and the halting rate can be balanced and become independent of time}, the trajectory is said to be in equilibrium with $\phi$.

The aforementioned notion of equilibrium should not be confused with the \emph{detailed balance}, which is the balance between the process \eqnref{register} and its reverse process \eqnref{reverse process}. More precisely, if the transition rates of \eqnref{register} and \eqnref{reverse process} are given by $\dot{P}$ and $\dot{P}_r$, respectively, the principle of detailed balance dictates that both are independent of $\tau$ and satisfy
\begin{equation}\label{detailed balance}
\frac{\dot{P}}{\dot{P}_r}=e^{-\beta\Delta E},
\end{equation}
where $1/\beta\equiv k_\mathrm{B}T$ is to be interpreted as the corresponding temperature. If the detailed balance is reachable, \emph{without employing the halting and reviving mechanisms}, the ratio of $\dot{P}$ to $\dot{P}_r$ is measured simply by the ratio of the number of registered detectors to the number of unregistered ones, i.e.,
\begin{equation}
\frac{\dot{P}}{\dot{P}_r} = \frac{N}{N_0-N},
\end{equation}
where $N$ is the number of registered detectors (averaged over probabilistic fluctuations in time). Because $\dot{P}$ and $\dot{P}_r$ are independent of $\tau$, the detailed balance entails the condition that the trajectory is in equilibrium with the background field. Conversely, however, it is not clear whether the condition of equilibrium must imply the detailed balance, although it does in many known examples (but see \footref{foot:counterexample}).
In the celebrated example of a uniformly accelerated detector in the Minkowski spacetime, the detailed balance relation can be shown to be satisfied from the Rindler observer perspective (see Secs.\ III.A.2 and III.A.4 of \cite{Crispino:2007eb} for more details.) This relies on the fact that the Minkowski vacuum is a thermal state of the right (left) Rindler modes if the left (right) Rindler modes are ignored (see \footref{foot:Unruh effect}). That is, from the Rindler observer perspective, we have
\begin{equation}\label{detailed balance 2}
\frac{\dot{P}}{\dot{P}_r} =\frac{\abs{\mathcal{A}}^2 n(\Delta E)}{\abs{\mathcal{A}_r}^2\left(1+n(\Delta E)\right)}
=\frac{n(\Delta E)}{1+n(\Delta E)}
=e^{-\beta\Delta E},
\end{equation}
where $\mathcal{A}$ and $\mathcal{A}_r$ are the amplitudes measured by the Rindler observer for the process \eqnref{register} and its reverse process \eqnref{reverse process}, respectively, which are complex conjugate to each other because of unitarity, and
\begin{equation}
n(\omega) = \frac{1}{e^{\beta\omega}-1}
\end{equation}
is the Rindler particle number density for the thermal state with the Unruh temperature given by \eqnref{temperature}. The factor $n(\Delta E)$ in the numerator in \eqnref{detailed balance 2} is associated with the induced absorption of a Rindler particle from the thermal bath, and the factor $1+n(\Delta E)$ in the denominator is associated with the spontaneous and induced emissions of a Rindler particle to the thermal bath.\footnote{See the discussion after \eqnref{tau and xi 2} for the fact that emission (absorption) of a Minkowski particle is equivalent to absorption (emission) of a Rindler particle.}

For the detector with a constant acceleration in noncompact directions (\secref{sec:acceleration in noncompact directions}), the comoving observer can still be viewed as a Rindler observer and thus the same argument for \eqnref{detailed balance 2} can be carried over in the obvious way. Therefore, the detailed balance is satisfied with the same temperature given by \eqnref{temperature}. On the other hand, for the detector with a constant acceleration in the compact directions (\secref{sec:acceleration in compact direction}), the condition of equilibrium is not satisfied and one cannot make sense of the detailed balance. Finally, for the detector with a constant velocity (\secref{sec:constant velocity}), the ratio of $\dot{P}$ to $\dot{P}_r$ can be easily computed from the perspective of a non-moving observer. That is,
\begin{equation}\label{detailed balance 3}
\frac{\dot{P}}{\dot{P}_r} =\frac{\abs{\mathcal{A}}^2 \left(1+n(\Delta E)\right)}{\abs{\mathcal{A}_r}^2n(\Delta E)},
\end{equation}
where $\mathcal{A}$ and $\mathcal{A}_r$ are the amplitudes measured by the non-moving observer, which are complex conjugate to each other, and
\begin{equation}
n(\omega_\mathbf{k}) :=  \bra{0_L} a^\dagger_\mathbf{k} a_\mathbf{k} \ket{0_L} = 0
\end{equation}
is the particle number density for the vacuum state $\ket{0_L}$. The factor $1+n(\Delta E)$ is associated with the spontaneous and induced emissions of a particle to $\ket{0_L}$, and the factor $n(\Delta E)$ is associated with the induced absorption of a particle from $\ket{0_L}$. For $\Delta E<0$, we have $\abs{\mathcal{A}}^2=\abs{\mathcal{A}_r}^2\neq0$ by \eqnref{rate for constant velocity b}, and consequently \eqnref{detailed balance 3} yields $\dot{P}/\dot{P}_r=\infty$. Therefore, the detailed balance is satisfied in the trivial way corresponding to the zero temperature $T=0$ (i.e., $\beta=\infty)$. For $\Delta E>0$, we have $\abs{\mathcal{A}}^2=\abs{\mathcal{A}_r}^2=0$ by \eqnref{rate for constant velocity a}, and the ratio \eqnref{detailed balance 3} is ill defined.\footnote{\label{foot:counterexample}This can be viewed as a trivial counterexample against the proposition that the condition of equilibrium always implies the detailed balance.}

Furthermore, the detailed balance should not be confused with the equilibrium between the transition rates of excitation and de-excitation. In fact, \eqnref{register a} and \eqnref{register b} are two \emph{independent} processes and their transition rates are measured independently by two different devices. Therefore, one cannot make sense of the equilibrium between them even when they share the same $\abs{\Delta E}$. This confusion arises partly because the words \emph{excitation} and \emph{de-excitation} are used in the literature to refer to the two different dichotomies. On the one hand, they refer to the dichotomy between \eqnref{register a} and \eqnref{register b} (as, e.g., in \cite{Louko:2006zv,Satz:2006kb,Louko:2007mu} as well as in this paper), but on the other hand,  they refer to the dichotomy between \eqnref{register} and \eqnref{reverse process} (as, e.g., in \cite{Crispino:2007eb}). In the latter terminological usage, \eqnref{register a} and \eqnref{register b} are both described as \emph{excitation} despite the fact that the detector is \emph{de-excited} from $\ket{E_0}$ to $\ket{E}$ in \eqnref{register b}. Perhaps \emph{registerization} (of a signal) and \emph{de-registerization} are the better words for the dichotomy between \eqnref{register} and \eqnref{reverse process}.

\section{Details for Section \ref{sec:with switching}}\label{app:detail 1}
Here, we present the detailed derivation from \eqnref{response function} to \eqnref{F final form}, following the same strategies in \cite{Louko:2006zv,Satz:2006kb} (also see \cite{Louko:2007mu} for the related techniques).

By the change of variables: $u=\tau$, $s=\tau-\tau'$ for $\tau'<\tau$ and $u=\tau'$, $s=\tau'-\tau$ for $\tau'>\tau$, the response function \eqnref{response function} can be recast as
\begin{equation}\label{F1}
F(\Delta E) = 2\, \mathrm{Re} \int_\infty^\infty du\, \chi(u)
\int_0^\infty ds\, \chi(u-s)\, e^{i\Delta Es} G^+_L(u,u-s).
\end{equation}
By \eqnref{GL+ 2} and \eqnref{G+}, we have
\begin{equation}
G^+_L(u,u-s) = -\frac{1}{4\pi^2}\sum_{n=-\infty}^\infty
\frac{1}{(\Delta t-i\epsilon)^2-\Delta \mathbf{x}_n^2},
\end{equation}
where $\Delta t$ and $\Delta\mathbf{x}_n^2$ are defined in \eqnref{Dt and Dxn}.
Eq.~\eqnref{F1} then leads to
\begin{eqnarray}\label{F2}
F(\Delta E) &=& \sum_{n=-\infty}^\infty
\lim_{\epsilon\rightarrow0}\frac{1}{2\pi^2}
\int_\infty^\infty du\, \chi(u)\int_0^\infty ds\, \chi(u-s)\nonumber\\
&&\qquad \mbox{} \times
\left[
\frac{\cos(\Delta Es)\left(\Delta x_n^2+\epsilon^2\right)
-2\sin(\Delta Es)\epsilon\Delta t}
{\left(\Delta x_n^2+\epsilon^2\right)^2+4\epsilon^2\Delta t^2}
\right],
\end{eqnarray}
where $\Delta x_n^2$ and $\Delta x_{n=0}^2\equiv(\Delta x)^2$ are defined in \eqnref{Dx and Dxn}.

We separate the integral over $s$ for a given $n$ in \eqnref{F2} into four parts: $I_>^\mathrm{even}$, $I_>^\mathrm{odd}$, $I_<^\mathrm{even}$, and $I_<^\mathrm{odd}$, where $>$ and $<$ denote the integrals over the intervals $[\sqrt{\epsilon},\infty)$ and $[0,\sqrt{\epsilon}]$, respectively, while ``even'' and ``odd'' denote the terms even and odd in $\Delta E$. The part $I_>^\mathrm{even}$ can be recast as
\begin{eqnarray}\label{I>even}
I_>^\mathrm{even}
&:=&
\int_{\sqrt{\epsilon}}^\infty ds\, \chi(u-s)\
\frac{\cos(\Delta Es)\left(\Delta x_n^2+\epsilon^2\right)}
{\left(\Delta x_n^2+\epsilon^2\right)^2+4\epsilon^2\Delta t^2}\nonumber\\
&=&
\int_{\sqrt{\epsilon}}^\infty ds\, \chi(u-s)\, \frac{\cos(\Delta E s)}{\Delta x_n^2}
\left[1-\frac{\epsilon^2}{\Delta x_n^2}
\frac{1+4\frac{\Delta t^2}{\Delta x_n^2}+\frac{\epsilon^2}{\Delta x_n^2}}
{\left(1+\frac{\epsilon^2}{\Delta x_n^2}\right)^2+4\epsilon^2\frac{\Delta t^2}{(\Delta x_n^2)^2}}\right].
\end{eqnarray}
To proceed, we need to know the small $s$ expansions for various variables. They can be obtained quite straightforwardly. Take the least trivial case of $(\Delta x)^2$ as an example. It follows from \eqnref{Dx square} that
\begin{subequations}
\begin{eqnarray}
(\Delta x)^2 &=& \big(x^\mu(u)-x^\mu(u-s)\big)\big(x_\mu(u)-x_\mu(u-s)\big),\\
\frac{d}{ds}(\Delta x)^2 &=& 2 \big(x^\mu(u)-x^\mu(u-s)\big)\dot{x}_\mu(u-s),\\
\frac{d^2}{ds^2}(\Delta x)^2 &=& 2 \dot{x}^\mu(u-s)\dot{x}_\mu(u-s)
-2\big(x^\mu(u)-x^\mu(u-s)\big)\ddot{x}_\mu(u-s) \nonumber\\
&=& -2 -2\big(x^\mu(u)-x^\mu(u-s)\big)\ddot{x}_\mu(u-s),\\
\frac{d^3}{ds^3}(\Delta x)^2 &=& -2 \dot{x}^\mu(u-s)\ddot{x}_\mu(u-s)
+2\big(x^\mu(u)-x^\mu(u-s)\big)\dddot{x}_\mu(u-s) \nonumber\\
&=& 2\big(x^\mu(u)-x^\mu(u-s)\big)\dddot{x}_\mu(u-s),\\
\frac{d^4}{ds^4}(\Delta x)^2 &=& 2 \dot{x}^\mu(u-s)\dddot{x}_\mu(u-s)
-2\big(x^\mu(u)-x^\mu(u-s)\big)\ddddot{x}_\mu(u-s) \nonumber\\
&=& -2\ddot{x}^\mu(u-s)\ddot{x}_\mu(u-s)-2\big(x^\mu(u)-x^\mu(u-s)\big)\ddddot{x}_\mu(u-s),
\end{eqnarray}
\end{subequations}
where we have used $\dot{x}_\mu\dot{x}^\mu=-1$, $\ddot{x}_\mu\dot{x}^\mu=0$, and $\dddot{x}_\mu\dot{x}^\mu=-\ddot{x}_\mu\ddot{x}^\mu$ and the over-dot denotes the derivative with respect to $u$.
Consequently, the Taylor series in terms of small $s$ gives
\begin{equation}
(\Delta x)^2 = \sum_{n=0}^\infty \frac{1}{n!}\frac{d^{(n)}(\Delta x)^2}{ds^n}\Big|_{s=0} s^n
=-s^2 -\frac{1}{12}\ddot{x}^2 s^4 + O(s^5).
\end{equation}
To sum up, we have $\Delta x^\mu = \dot{x}^\mu s+O(s^2)$, which follows $\Delta t = O(s)$, $\Delta z = O(s)$, $(\Delta x)^2 = O(s^2)$, and $\Delta x_n^2 = (nL)^2 + nL\cdot O(s) +O(s^2)$.
Here, we assume an additional condition that $\Delta x_n^2$ is everywhere nonvanishing except $s=0$ (we will come back to this additional condition shortly). Provided with this additional condition, the quantities $\abs{\Delta t^2/\Delta x_n^2}$ and $\abs{\epsilon/\Delta x_n^2}$ are both bounded by some constants independent of $\epsilon$ over the intersection of the compact support of $\chi(u-s)$ and the interval $[\sqrt{\epsilon},\infty)$.\footnote{\label{foot:bounds}Let $s\in[A,B]$ denote the compact support of $\chi(u-s)$. The intersection $[A,B]\cap[\sqrt{\epsilon},\infty)$ (if not empty) is either $[A,B]$, or $[\sqrt{\epsilon},B]$. In the former case, because $[A,B]$ is compact, the upper bound of $\abs{\Delta t^2/\Delta x_n^2}$ and the lower bound of $\abs{\Delta x_n^2}$ occur at $s=A$, $s=B$, or some point in between where the corresponding derivative vanishes. In all three situations, the bounds are independent of $\epsilon$. Furthermore, the lower bound of $\abs{\Delta x_n^2}$ is nonzero, since we have assumed $\Delta x_n^2\neq0$ except $s=0$. Consequently, $\abs{t^2/\Delta x_n^2}$ and $\abs{\epsilon/\Delta x_n^2}<\abs{A/\Delta x_n^2}$ are both bounded by some constants independent of $\epsilon$. In the latter case, the upper bound of $\abs{\Delta t/\Delta x_n^2}$ and the lower bound of $\abs{\Delta x_n^2}$ might occur at $s=\sqrt{\epsilon}$ and thus depend on $\epsilon$. Nevertheless, the $s\rightarrow\sqrt{\epsilon}$ behaviors given by the small $s$ expansions yield $\abs{\Delta t^2/\Delta x_n^2}=O(s^2)/\left((nL)^2 + nL\cdot O(s) +O(s^2)\right) \leq O(s^2)/O(s^2)=O(s^0)$ and $\abs{\epsilon/\Delta x_n^2}\leq\epsilon/O(s^2)\rightarrow O(\epsilon^0)$. Therefore, in both cases, $\abs{\Delta t^2/\Delta x_n^2}$ and $\abs{\epsilon/\Delta x_n^2}$ are bounded by some constants independent of $\epsilon$.}. Consequently, the absolute value of the second part of \eqnref{I>even} in the $\epsilon\rightarrow0$ limit yields
\begin{eqnarray}\label{error piece}
&&
\abs{
\int_{\sqrt{\epsilon}}^\infty ds\, \chi(u-s)\cos(\Delta E s)\,
\frac{\epsilon^2}{(\Delta x_n^2)^2}\,
\frac{1+4\frac{\Delta t^2}{\Delta x_n^2}+\epsilon\,\frac{\epsilon}{\Delta x_n^2}}
{\left(1+\epsilon\,\frac{\epsilon}{\Delta x_n^2}\right)^2+4\epsilon\,\frac{\epsilon}{\Delta x_n^2}\frac{\Delta t^2}{\Delta x_n^2}}
}\nonumber\\
&\mathop{\longrightarrow}\limits_{\epsilon\rightarrow0}&
\abs{
\int_{\sqrt{\epsilon}}^\infty ds\, \chi(u-s)\cos(\Delta E s)\,
\frac{\epsilon^2}{(\Delta x_n^2)^2}\,
\left(1+4\frac{\Delta t^2}{\Delta x_n^2}\right)
}\nonumber\\
&\leq&
\int_{\sqrt{\epsilon}}^B ds\,
\abs{\frac{\epsilon^2}{(\Delta x_n^2)^2}}(1+4C_1)
= C_2\int_{\sqrt{\epsilon}}^B ds\,
\abs{\frac{\epsilon^2}{(\Delta x_n^2)^2}},
\end{eqnarray}
where $C_1$ is some constant such that $\abs{\Delta t^2/\Delta x_n^2}\leq C_1$, $C_2:=1+4C_1$, and $B$ is the upper boundary of the compact support of $\chi(u-s)$.
As discussed in \footref{foot:bounds}, if the lower bound of $\abs{\Delta x_n^2}$ does not occur at $s=\sqrt{\epsilon}$, we have $\abs{\Delta x_n^2}\geq C_3$ for some constant $C_3\neq0$, which is independent of $\epsilon$. Consequently, \eqnref{error piece} leads to
\begin{equation}\label{error piece 2}
C_2\int_{\sqrt{\epsilon}}^B ds\,
\abs{\frac{\epsilon^2}{(\Delta x_n^2)^2}}
\leq \frac{C_2\epsilon^2}{C_3^2}\int_{\sqrt{\epsilon}}^B ds
\leq\frac{C_2\epsilon^2(B-\sqrt{\epsilon})}{C_3^2}
=O(\epsilon^2).
\end{equation}
On the other hand, if the lower bound of $\abs{\Delta x_n^2}$ occurs at $s=\sqrt{\epsilon}$, let $M\in(\sqrt{\epsilon},B)$ be the turning point of $\abs{\Delta x_n^2}$ if the turning point exists and let $M=B$ if it does not. That is, $\abs{\Delta x_n^2}$ is monotonically increasing from $s=\sqrt{\epsilon}$ until $s=M$. The turning point, if it exists, is located at the point where the derivative of $\Delta x_n^2$ with respect to $s$ vanishes and thus $M$ is independent of $\epsilon$. Within the compact interval $[M,B]$, we have $\abs{\Delta x_n^2}\geq C_4$ for some lower bound constant $C_4\neq0$, which is independent of $\epsilon$. Consequently, \eqnref{error piece} leads to
\begin{eqnarray}\label{error piece 3}
&&C_2\int_{\sqrt{\epsilon}}^B ds\,
\abs{\frac{\epsilon^2}{(\Delta x_n^2)^2}}
\leq
C_2\int_{\sqrt{\epsilon}}^M ds \frac{\epsilon^2}{(as^2)^2}
+C_2\int_{M}^B ds\,\frac{\epsilon^2}{C_4^2} \nonumber\\
&=& \left.-\frac{C_2\epsilon^2}{3a^2s^3}\right|_{s=\sqrt{\epsilon}}^M
+\frac{C_2\epsilon^2}{C_4^2} (B-M)
= O(\sqrt{\epsilon}) + O(\epsilon^2),
\end{eqnarray}
where the small $s$ expansion $\Delta x_n^2 = (nL)^2 + nL\cdot O(s) +O(s^2)$ allows us to apply the inequality $\abs{\Delta x_n^2}\geq as^2$ with some constant $a$ in the interval $[\sqrt{\epsilon},M]$, since $\abs{\Delta x_n^2}$ is monotonically increasing within $[\sqrt{\epsilon},M]$.
It follows from \eqnref{error piece 2} and \eqnref{error piece 3} that the second part of \eqnref{I>even} is of order $O(\sqrt{\epsilon})$ at worst in the $\epsilon\rightarrow0$ limit. Therefore, we have
\begin{equation}\label{I>even 2}
I_>^\mathrm{even}
=
\int_{\sqrt{\epsilon}}^\infty ds\, \chi(u-s)\, \frac{\cos(\Delta E s)}{\Delta x_n^2}
+ O(\sqrt{\epsilon}).
\end{equation}

Similarly, The part $I_>^\mathrm{odd}$ can be recast as
\begin{eqnarray}\label{I>odd}
I_>^\mathrm{odd}
&:=&
\int_{\sqrt{\epsilon}}^\infty ds\, \chi(u-s)\
\frac{-2\sin(\Delta Es)\epsilon\Delta t}
{\left(\Delta x_n^2+\epsilon^2\right)^2+4\epsilon^2\Delta t^2}\nonumber\\
&=&
\int_{\sqrt{\epsilon}}^\infty ds\, \chi(u-s)\, \frac{\sin(\Delta E s)}{(\Delta x_n^2)^2}
\frac{-2\epsilon\Delta t}
{\left(1+\frac{\epsilon^2}{\Delta x_n^2}\right)^2+4\epsilon^2\frac{\Delta t^2}{(\Delta x_n^2)^2}}.
\end{eqnarray}
Following the same argument for the second part of $I_>^\mathrm{even}$, we have
\begin{eqnarray}
\abs{I_>^\mathrm{odd}} &\leq&
2\int_{\sqrt{\epsilon}}^\infty ds\, \chi(u-s)\,
\frac{\epsilon\abs{\sin(\Delta E s)}}{\abs{\Delta x_n^2}^{3/2}}
\sqrt{\abs{\frac{\Delta t^2}{\Delta x_n^2}}}\
\frac{1}{\left(1+\epsilon\,\frac{\epsilon}{\Delta x_n^2}\right)^2+4\epsilon\,\frac{\epsilon}{\Delta x_n^2}\frac{\Delta t^2}{\Delta x_n^2}}\nonumber\\
&\mathop{\longrightarrow}\limits_{\epsilon\rightarrow0}&
2\int_{\sqrt{\epsilon}}^\infty ds\, \chi(u-s)\,
\frac{\epsilon\abs{\sin(\Delta E s)}}{\abs{\Delta x_n^2}^{3/2}}
\sqrt{\abs{\frac{\Delta t^2}{\Delta x_n^2}}}\nonumber\\
&\leq&
2C_1\int_{\sqrt{\epsilon}}^B ds\,
\frac{\epsilon\abs{\sin(\Delta E s)}}{\abs{\Delta x_n^2}^{3/2}}.
\end{eqnarray}
If the lower bound of $\abs{\Delta x_n^2}$ does not occur at $s=\sqrt{\epsilon}$, we have
\begin{equation}
\abs{I_>^\mathrm{odd}}
\leq
2C_1\int_{\sqrt{\epsilon}}^B ds\,\frac{\epsilon}{C_3^{3/2}}
= \frac{2C_1\epsilon(B-\sqrt{\epsilon})}{C_3^{3/2}}
= O(\epsilon).
\end{equation}
If the lower bound of $\abs{\Delta x_n^2}$ occurs at $s=\sqrt{\epsilon}$, we have
\begin{eqnarray}
\abs{I_>^\mathrm{odd}}
&\leq&
2C_1\int_{\sqrt{\epsilon}}^M ds\,\frac{\epsilon\abs{\Delta E}s}{(as^2)^{3/2}}
+2C_1\int_M^B ds\,\frac{\epsilon}{C_3^{3/2}} \nonumber\\
&=& \left.-\frac{2C_1\epsilon\abs{\Delta E}}{a^{3/2}s}\right|_{s=\sqrt{\epsilon}}^M
+\frac{2C_1\epsilon(B-\sqrt{\epsilon})}{C_3^{3/2}}
= O(\sqrt{\epsilon}) + O(\epsilon),
\end{eqnarray}
where we have used $\sin x\leq x$ for $x\geq0$. Therefore, we have
\begin{equation}\label{I>odd 2}
I_>^\mathrm{odd}
=
O(\sqrt{\epsilon}).
\end{equation}

The aforementioned additional condition is satisfied in the case of the ordinary Minkowski spacetime (where only the $n=0$ term of $\Delta x_n^2$ survives), but it can be violated in general as in some examples in this paper. In other words, $1/\Delta x_n^2$ might become singular at some points other than $s=0$. Nevertheless, \eqnref{I>even 2} and \eqnref{I>odd 2} can still be taken as valid, if the switching function $\chi(u-s)$ is turned off before the integrating variable $s$ encounters any singularities of $1/\Delta x_n^2$.

Next, let us study the remaining half part $I_<^\mathrm{even}$ and $I_<^\mathrm{odd}$:
\begin{equation}\label{I<}
I_<^\mathrm{even} + I_<^\mathrm{odd} :=
\int_0^{\sqrt{\epsilon}} ds\, \chi(u-s)
\left[
\frac{\cos(\Delta Es)\left(\Delta x_n^2+\epsilon^2\right)
-2\sin(\Delta Es)\epsilon\Delta t}
{\left(\Delta x_n^2+\epsilon^2\right)^2+4\epsilon^2\Delta t^2}
\right].
\end{equation}
For $n\neq0$, we have $\Delta x_n^2 = (nL)^2 + O(s)$ and $\Delta t=O(s)$ as $s\rightarrow0$. It follows that the absolute value of the integrand of $I_<^\mathrm{even}$ is of order $O(s^0/(nL)^2)$ and that of $I_<^\mathrm{odd}$ is of order $O(\epsilon s^2/(nL)^4)$, rendering $I_<^\mathrm{even}$ to be of order $O(\sqrt{\epsilon})$ and $I_<^\mathrm{odd}$ of order $O(\epsilon^{5/2})$. For $n=0$, as we have $\Delta x_{n=0}^2\equiv (\Delta x)^2 = O(s^2)$ and the integrand in \eqnref{I<} grows to infinity as $s\rightarrow0$, the result is expected to be nonzero in the limit $\epsilon\rightarrow0$. It is quite involved to compute the result of the $n=0$ part, but it is exactly the same as that calculated in \cite{Satz:2006kb} and we simply cite the result:\footnote{The derivation can be found in Sec.~3 of \cite{Satz:2006kb} and some related details are given in \cite{Louko:2006zv}, especially Appendix A therein. The first step is to make the change of variable $s=\epsilon x$ and recast $I_<^\mathrm{even}$ and $I_<^\mathrm{odd}$ as
\begin{eqnarray*}
I_<^\mathrm{even} &=& \frac{1}{\epsilon}\int_0^{1/\sqrt{\epsilon}}
dx\,
\frac{\left(\chi-\dot{\chi}\epsilon x\right)\left(1-x^2\right)}
{1+x^4+2x^2\left(2\dot{t}^2-1\right)}
\left[1
+\frac{4\dot{t}\,\ddot{t}\,\epsilon x^3}
{1+x^4+2x^2\left(2\dot{t}^2-1\right)}
\right]
+O(\sqrt{\epsilon}),\\
I_<^\mathrm{odd} &=& -\int_0^{1/\sqrt{\epsilon}}
dx\,
\frac{2\chi\,\Delta E\, \dot{t}\,x^2}
{1+x^4+2x^2\left(2\dot{t}^2-1\right)}
+O(\sqrt{\epsilon}).
\end{eqnarray*}
Not that the corresponding formula (3.4) in \cite{Satz:2006kb} has a typo.}
\begin{eqnarray}\label{I< n=0}
I_<^\mathrm{even} + I_<^\mathrm{odd}
&=& \frac{\chi}{\sqrt{\epsilon}} -\dot{\chi}\ln\sqrt{\epsilon} -\frac{\pi\Delta E\chi}{2}
+\frac{\dot{\chi}\,\dot{t}}{(\dot{t}^2-1)^{1/2}}\ln\big(\dot{t}-(\dot{t}^2-1)^{1/2}\big)
\nonumber\\
&&\mbox{}
-\frac{\dot{\chi}\,\ddot{t}}{2(\dot{t}^2-1)^{3/2}}
\left[\dot{t}(\dot{t}^2-1)^{1/2}+\ln\big(\dot{t}-(\dot{t}^2-1)^{1/2}\big)\right]
+O(\sqrt{\epsilon}).
\end{eqnarray}
The term $-\dot{\chi}\ln\sqrt{\epsilon}$ in the above vanishes via integration by parts when plugged into \eqnref{F2}, since $\chi(u)$ is smooth and becomes zero before and after the interaction. Furthermore, because
\begin{equation}
\frac{d}{du}\left[\frac{\dot{t}}{(\dot{t}^2-1)^{1/2}}\ln\big(\dot{t}-(\dot{t}^2-1)^{1/2}\big)\right]
=-\frac{\ddot{t}}{2(\dot{t}^2-1)^{3/2}}
\left[\dot{t}(\dot{t}^2-1)^{1/2}+\ln\big(\dot{t}-(\dot{t}^2-1)^{1/2}\big)\right],
\end{equation}
the last two terms in \eqnref{I< n=0} cancel out via integration by parts in \eqnref{F2}. To sum up, we have
\begin{subequations}\label{I< 2}
\begin{eqnarray}
I_<^\mathrm{even} + I_<^\mathrm{odd}
&=&\frac{\chi}{\sqrt{\epsilon}} -\frac{\pi\Delta E\chi}{2} + O(\sqrt{\epsilon}),
\qquad \text{for } n=0, \\
&=& O(\sqrt{\epsilon}),
\qquad \text{for } n\neq0.
\end{eqnarray}
\end{subequations}

Provided that the aforementioned additional condition is satisfied or that the switching function $\chi(u-s)$ is turned off before $s$ hits any nonzero singularities of $1/\Delta x_n^2$, putting \eqnref{I>even 2}, \eqnref{I>odd 2}, and \eqnref{I< 2} altogether into \eqnref{F2} then gives
\begin{eqnarray}
&&F(\Delta E) \\
&=& \lim_{\epsilon\rightarrow0}
\frac{1}{2\pi^2} \int_{-\infty}^\infty du\, \chi(u)
\left[
-\frac{\pi\Delta E\chi(u)}{2} + \int_{\sqrt{\epsilon}}^\infty ds
\left(\chi(u-s)\cos(\Delta E s)\sum_{n=-\infty}^\infty\frac{1}{\Delta x_n^2}
+\frac{\chi(u)}{s^2}\right)
\right], \nonumber
\end{eqnarray}
where $\chi(u)/\sqrt{\epsilon}$ has been replaced by $\int_{\sqrt{\epsilon}}^\infty ds \frac{\chi(u)}{s^2}$.
Removing the regularization by taking the $\epsilon\rightarrow0$ limit then yields \eqnref{F final form}, where we have added and subtracted a term $\chi(u-s)/s^2$ for the reason to facilitate the analysis of the sharp switching limit studied in \secref{sec:sharp switching limit}.

\section{Details for \eqnref{GL+ a in z} and \eqnref{GL+ a in x}}\label{app:detail 2}
Here we present the detailed derivation from \eqnref{GL+ a in z original} to \eqnref{GL+ a in z}. The similar trick is also applied in \eqnref{GL+ a in x}.

The denominator of each summand in \eqnref{GL+ a in z original} can be recast as
\begin{subequations}\label{denominator}
\begin{eqnarray}
&&\left(\alpha\sinh\frac{\tau}{\alpha}-\alpha\sinh\frac{\tau'}{\alpha}-i\epsilon\right)^2
-\left(\alpha\cosh\frac{\tau}{\alpha}-\alpha\cosh\frac{\tau'}{\alpha}-nL\right)^2 \nonumber\\
&=& -\alpha^2\left(\cosh^2\frac{\tau}{\alpha}-\sinh^2\frac{\tau}{\alpha} +\cosh^2\frac{\tau'}{\alpha}-\sinh^2\frac{\tau'}{\alpha}\right)
+2\alpha^2\left(\cosh\frac{\tau}{\alpha}\cosh\frac{\tau'}{\alpha} -\sinh\frac{\tau}{\alpha}\sinh\frac{\tau'}{\alpha}\right) \nonumber\\
&&\mbox{} -2i\epsilon\alpha\left(\sinh\frac{\tau}{\alpha}-\sinh\frac{\tau}{\alpha}'\right)
+2nL\alpha\left(\cosh\frac{\tau}{\alpha}-\cosh\frac{\tau}{\alpha}'\right)
-n^2L^2 + O(\epsilon^2) \nonumber\\
&=& -2\alpha^2 +2\alpha^2\cosh\frac{\tau-\tau'}{\alpha}
-4i\epsilon\alpha\cosh\frac{\tau+\tau'}{2\alpha}\sinh\frac{\tau-\tau'}{2\alpha}
+4nL\alpha\sinh\frac{\tau+\tau'}{2\alpha}\sinh\frac{\tau-\tau'}{2\alpha} \nonumber\\
&& \mbox{} -n^2L^2 +O(\epsilon^2) \nonumber\\
\label{denominator a}
&=& 4\alpha^2\sinh^2\frac{\tau-\tau'}{2\alpha}
-4i\epsilon\alpha\sinh\frac{\tau-\tau'}{2\alpha}
\left(\cosh\frac{\tau-\tau'}{2\alpha}\cosh\frac{\tau'}{\alpha} -\sinh\frac{\tau-\tau'}{2\alpha}\sinh\frac{\tau'}{\alpha}\right) \nonumber\\
&& \mbox{} +4nL\alpha\sinh\frac{\tau+\tau'}{2\alpha}\sinh\frac{\tau-\tau'}{2\alpha} -n^2L^2 +O(\epsilon^2)\\
\label{denominator b}
&=& 4\alpha^2\sinh^2\frac{\Delta\tau}{2\alpha}\left(1+O(\epsilon)\right)
-4i\epsilon\alpha\sinh\frac{\Delta\tau}{2\alpha}\cosh\frac{\Delta\tau}{2\alpha} \nonumber\\
&& \mbox{} +4nL\alpha\sinh\frac{\tau+\tau'}{2\alpha}\sinh\frac{\Delta\tau}{2\alpha} -n^2L^2
+O(\epsilon^2)\\
\label{denominator c}
&\approx& 4\alpha^2\sinh^2\left(\frac{\Delta\tau}{2\alpha}-\frac{i\epsilon}{2\alpha}\right)
+4nL\alpha\sinh\frac{\tau+\tau'}{2\alpha}\sinh\frac{\Delta\tau}{2\alpha} -n^2L^2,
\end{eqnarray}
\end{subequations}
where from \eqnref{denominator a} to \eqnref{denominator b} we have absorbed the positive factor $\cosh (\tau'/\alpha)$ into $\epsilon$ and from \eqnref{denominator b} to \eqnref{denominator c} we have used
\begin{eqnarray}
&&\sinh\left(\frac{\Delta\tau}{2\alpha}-\frac{i\epsilon}{2\alpha}\right)
= \sinh\frac{\Delta\tau}{2\alpha}\cosh\frac{i\epsilon}{2\alpha} -\cosh\frac{\Delta\tau}{2\alpha}\sinh\frac{i\epsilon}{2\alpha}\nonumber\\
&=& \sinh\frac{\Delta\tau}{2\alpha} -\frac{i\epsilon}{2\alpha}\cosh\frac{\Delta\tau}{2\alpha}
+O(\epsilon^2).
\end{eqnarray}


\begin{thebibliography}{99}

\bibitem{Debs:1996}
  T.~A.~Debs and M.~L.~G.~Redhead,
  ``The twin `paradox' and the conventionality of simultaneity,''
  Am.\ J.\ Phys.\ {\bf 64}, 384 (1996).

\bibitem{Brans:1973}
  C.~H.~Brans and D.~R.~Stewart
  ``Unaccelerated-Returning-Twin Paradox in Flat Space-Time,''
  Phys.\ Rev.\ D {\bf 8}, 1662 (1973).

\bibitem{Peters:1983as}
  P.~C.~Peters,
  ``Periodic boundary conditions in special relativity,''
  Am.\ J.\ Phys.\  {\bf 51}, 791 (1983).

\bibitem{Dray:1990}
 T.~Dray,
 ``The twin paradox revisited,''
 Am.\ J.\ Phys.\ {\bf 58}, 822 (1990).

\bibitem{Barrow:2001rj}
  J.~D.~Barrow and J.~J.~Levin,
  ``Twin paradox in compact spaces,''
  Phys.\ Rev.\ A {\bf 63}, 044104 (2001)
  [gr-qc/0101014].

\bibitem{Uzan:2000wp}
  J.~P.~Uzan, J.~P.~Luminet, R.~Lehoucq and P.~Peter,
  ``Twin paradox and space topology,''
  Eur.\ J.\ Phys.\  {\bf 23}, 277 (2002)
  [physics/0006039 [physics.class-ph]].

\bibitem{Bansal:2005ue}
  D.~Bansal, J.~Laing and A.~Sriharan,
  ``On the twin paradox in a universe with a compact dimension,''
  gr-qc/0503070

\bibitem{Hossain:2010eb}
  G.~M.~Hossain, V.~Husain and S.~S.~Seahra,
  ``Propagator in polymer quantum field theory,''
  Phys.\ Rev.\ D {\bf 82}, 124032 (2010)
  [arXiv:1007.5500 [gr-qc]].

\bibitem{Kajuri:2015oza}
  N.~Kajuri,
  ``Polymer quantization predicts radiation in inertial frames,''
  Class.\ Quant.\ Grav.\  {\bf 33}, no. 5, 055007 (2016)
  [arXiv:1508.00659 [gr-qc]].

\bibitem{Birrell:1982ix}
  N.~D.~Birrell and P.~C.~W.~Davies,
  {\it Quantum Fields in Curved Space},
  (Cambridge University Press, Cambridge, 1982).

\bibitem{Louko:2006zv}
  J.~Louko and A.~Satz,
  ``How often does the Unruh-DeWitt detector click? Regularisation by a spatial profile,''
  Class.\ Quant.\ Grav.\  {\bf 23}, 6321 (2006)
  [gr-qc/0606067].

\bibitem{Satz:2006kb}
  A.~Satz,
  ``Then again, how often does the Unruh-DeWitt detector click if we switch it carefully?,''
  Class.\ Quant.\ Grav.\  {\bf 24}, 1719 (2007)
  [gr-qc/0611067].

\bibitem{Louko:2007mu}
  J.~Louko and A.~Satz,
  ``Transition rate of the Unruh-DeWitt detector in curved spacetime,''
  Class.\ Quant.\ Grav.\  {\bf 25}, 055012 (2008)
  [arXiv:0710.5671 [gr-qc]].

\bibitem{Fulling:1972md}
  S.~A.~Fulling,
  ``Nonuniqueness of canonical field quantization in Riemannian space-time,''
  Phys.\ Rev.\ D {\bf 7}, 2850 (1973).

\bibitem{Glauber:1963fi}
  R.~J.~Glauber,
  ``The Quantum theory of optical coherence,''
  Phys.\ Rev.\  {\bf 130}, 2529 (1963).

\bibitem{Unruh:1976db}
  W.~G.~Unruh,
  ``Notes on black hole evaporation,''
  Phys.\ Rev.\ D {\bf 14}, 870 (1976).

\bibitem{Sanchez:1981xx}
  N.~G.~S\'{a}nchez,
  ``Quantum Detection on the Vacuum by Nonuniformly Accelerated Observers,''
  Phys.\ Lett.\  {\bf 105B}, 375 (1981).

\bibitem{DeWitt:1979}
  B.~S.~DeWitt,
  ``Quantum Gravity: The New Synthesis,'' in {\it General Relativity: an Einstein Centenary Survey}, eds.\ S.~W.~Hawking and W.~Israel (Cambridge University Press, Cambridge, 1979).

\bibitem{Wald:book}
 R.~M.~Wald,
 {\it Quantum Field Theory in Curved Spacetime and Black Hole Thermodynamics},
 (University of Chicago Press, Chicago, 1994).

\bibitem{Padmanabhan:2003gd}
  T.~Padmanabhan,
  ``Gravity and the thermodynamics of horizons,''
  Phys.\ Rept.\  {\bf 406}, 49 (2005)
  [gr-qc/0311036].

\bibitem{Crispino:2007eb}
  L.~C.~B.~Crispino, A.~Higuchi and G.~E.~A.~Matsas,
  ``The Unruh effect and its applications,''
  Rev.\ Mod.\ Phys.\  {\bf 80}, 787 (2008)
  [arXiv:0710.5373 [gr-qc]].

\bibitem{Davies:1989me}
  P.~C.~W.~Davies, Z.~X.~Liu and A.~C.~Ottewill,
  ``Particle Detectors in the Presence of Boundaries,''
  Class.\ Quant.\ Grav.\  {\bf 6}, 1041 (1989).

\bibitem{Schlicht:2003iy}
  S.~Schlicht,
  ``Considerations on the Unruh effect: Causality and regularization,''
  Class.\ Quant.\ Grav.\  {\bf 21}, 4647 (2004)
  [gr-qc/0306022].

\end{thebibliography}
\end{document}